\documentclass[aps,pra,twocolumn,showpacs,superscriptaddress,floatfix]{revtex4-1}
\usepackage[dvipdfmx]{graphicx}
\usepackage{natbib}
\usepackage{bm}
\usepackage[dvipsnames]{color} 
\usepackage{ulem}
\usepackage{amsmath}
\usepackage{amssymb}
\usepackage[
  colorlinks=true,
  citecolor=blue, 
  linkcolor=red,
  urlcolor=blue,
  setpagesize=false]{hyperref}

\newcommand{\bs}   {\boldsymbol}
\newcommand{\mb}   {\mathbf}

\newcommand{\mcal} {\mathcal}
\newcommand{\imag} {\mathrm{i}}
\newcommand{\dd}   {\mathrm{d}}
\newcommand{\e}    {\mathrm{e}}
\newcommand{\bra}  {\langle}
\newcommand{\ket}  {\rangle}
\newcommand{\up}   {\uparrow}
\newcommand{\dn}   {\downarrow}
\newcommand{\s}    {\sigma}
\newcommand{\w}    {\omega}
\newcommand{\eps}  {\epsilon}
\newcommand{\ofz} {(z)} 
\newcommand{\ofk} {(\vec{\mb{k}})} 
\newcommand{\ofkz} {(\vec{\mb{k}},z)} 
\newcommand{\veck} {\vec{\mb{k}}}
\newcommand{\veckt} {\vec{\tilde{{\mb{k}}}}}
\newcommand{\vecqs} {\vec{\mb{q}}_s}
\begin{document}

\title{Brillouin-zone integration scheme for many-body density of states:\\ 
  Tetrahedron method combined with cluster perturbation theory}

\author{K.~Seki}
\affiliation{Computational Condensed Matter Physics Laboratory, RIKEN, Wako, Saitama 351-0198, Japan}
\affiliation{Computational Materials Science Research Team, RIKEN Advanced Institute for Computational Science (AICS), Kobe, Hyogo 650-0047, Japan}
\author{S.~Yunoki}
\affiliation{Computational Condensed Matter Physics Laboratory, RIKEN, Wako, Saitama 351-0198, Japan}
\affiliation{Computational Materials Science Research Team, RIKEN Advanced Institute for Computational Science (AICS), Kobe, Hyogo 650-0047, Japan}
\affiliation{Computational Quantum Matter Research Team, RIKEN, Center for Emergent Matter Science (CEMS), Wako, Saitama 351-0198, Japan}

\begin{abstract} 
  By combining the tetrahedron method with the cluster perturbation theory (CPT), 
  we present an accurate method to numerically calculate the density of states of interacting fermions 
  without introducing the Lorentzian broadening parameter $\eta$ 
  or the numerical extrapolation of $\eta \to 0$.
  The method is conceptually based on the notion of the effective single-particle Hamiltonian which can be 
  subtracted in the Lehmann representation of the single-particle Green's function within the CPT. 
  Indeed, we show the general correspondence between the self-energy and 
  the effective single-particle Hamiltonian 
  which describes exactly the single-particle excitation energies of interacting fermions. 
  The detailed formalism is provided for two-dimensional multi-orbital systems and   
  a benchmark calculation is performed for the two-dimensional single-band Hubbard model. 
  The method can be adapted straightforwardly 
  to symmetry broken states, three-dimensional systems, and finite-temperature calculations.    
\end{abstract}

\pacs{
  71.10.-w,  
  05.30.Fk,  
  71.10.Fd   
}

\date{\today}
\maketitle

\section{Introduction}\label{sec:intro}

One of the important quantities in strongly correlated 
fermion systems is the single-particle excitation gap because  
the single-particle excitation gap plays the role of an ``order parameter'' 
in, e.g., the half-filled Hubbard models, which 
distinguishes the metallic state from the Mott insulating state 
in the absence of a long-range order. 
Theoretically, the single-particle excitation gap is usually estimated 
from the density of states near the Fermi energy. 
To calculate the the single-particle excitations, including the density of states, 
in strongly correlated systems, 
the quantum cluster approaches~\cite{Maier2005} are very often employed. 
These approaches require a numerical method (a solver) 
to treat many-body problems 
within small clusters. 
The numerically exact diagonalization is one of the methods 
which allow us to directly calculate the single-particle Green's functions at 
an arbitrary complex frequency~\cite{Dagotto1994,Weisse}. 

Even with the exact diagonalization, when the single-particle excitations 
are calculated, a finite imaginary $\eta$ of the complex frequency 
$z=\w+\imag \eta$ has to be introduced 
to avoid the poles of the single-particle Green's function 
lying on the real frequency $\w$ axis. 
Here, $\eta$ corresponds to the half width at half maximum of the Lorentzian broadening 
for a delta-function peak [see Eq.~(\ref{eq:dos})]. 
Since the density of states is  
provided as a sum of delta-function peaks, 
one has to calculate the density of states using several values of $\eta$ and 
extrapolate the results to $\eta \to 0$ for all frequencies $\w$, e.g., close to the chemical potential. 
The numerical extrapolation of $\eta \to 0$ is indeed valuable to 
examine fine structures of the density of states~\cite{Sahebsara2008,Yamada2015}.   
However, it is often technically cumbersome because appropriate values of $\eta$ 
have to be chosen to obtain reasonable, i.e., sharp enough and non-negative, density of states.

The single-particle excitation gap of interacting fermions can be evaluated not only from 
a direct calculation of the single-particle excitation energies~\cite{Laubach2015}  
but also from a jump of the chemical potential as a function of particle density~\cite{Seki2011,Seki,Hassan2013,Ebato2015}. 
However, the development of a theoretical method which can resolve fine structures of the many-body 
density of states is still highly valuable, as it allows us to make a direct comparison with 
experiments. For example, 
the scanning tunneling spectroscopy and microscopy (STS/STM) 
can prove the surface density of states for strongly correlated materials 
with high resolution~\cite{Balatsky2006,Fischer2007,Das2014}. 
  
In this paper,
we present a method to numerically calculate the density of states of interacting fermions 
by combining the tetrahedron method~\cite{Rath1975,Blochl1994}, widely used for noninteracting systems,  
with the cluster perturbation theory (CPT)~\cite{Senechal2000,Senechal2002,Senechal2012} 
in the Lehmann representation~\cite{Zacher2002,Aichhorn2006}. 
This method allows us to calculate the density of states of 
interacting fermions without introducing the Lorentzian broadening parameter $\eta$. 
In deriving the formalism, we emphasize the notion of the effective single-particle Hamiltonian 
for the single-particle excitations of interacting fermions, 
which is the conceptual basis of 
the method and bridges the gap between 
the single-particle theory and 
the single-particle excitations of interacting fermions in the many-body theory.

The rest of this paper is organized as follows.  
After briefly describing the basic formulation of the CPT in Sec.~\ref{sec:cpt},  
the Lehmann representation of the single-particle 
Green's function is introduced and the tetrahedron method for 
two-dimensional (2D) interacting fermions is described in Sec.~\ref{sec:formalism}.  
The effective single-particle Hamiltonian 
for the single-particle excitations of interacting fermions is also discussed. 
To demonstrate the method, a benchmark calculation is performed in Sec.~\ref{sec:benchmark}. 
Several remarks are made in Sec.~\ref{sec:discussion} to extend the method 
to symmetry broken states, three-dimensional (3D) systems, 
and finite-temperature calculations, before summarizing the paper 
in Sec.~\ref{sec:summary}.  
In order to ensure the notion of 
the effective single-particle Hamiltonian for the single-particle excitations of interacting fermions, 
the general correspondence between the self-energy and the effective single-particle Hamiltonian 
is established in appendix~\ref{appendix}. 
An additional technical detail to save the computational cost is provided in appendix~\ref{appendixB}.

\section{Cluster perturbation theory}\label{sec:cpt}

The CPT~\cite{Senechal2000,Senechal2002,Senechal2012} assumes that 
the Hamiltonian $\hat{H}$ on the infinite lattice can be divided into 
two parts, defined on a set of identical, disconnected finite-size clusters 
which cover all sites of the infinite lattice, i.e., 
\begin{equation}
\hat{H} = \sum_{I} \hat{H}_{\rm c} (\vec{\mb{R}}_I)
+ \sum_{\bra I,J \ket} \hat{\mcal{T}}(\vec{\mb{R}}_I - \vec{\mb{R}}_J ),
\end{equation} 
where $\hat H_{\rm c}(\vec{\mb{R}}_I)$ represents the Hamiltonian of the $I$-th finite-size cluster 
located at $\vec{\mb{R}}_I$, and $\langle I, J\rangle$ denotes a pair of 
different clusters at $\vec{\mb{R}}_I$ and $\vec{\mb{R}}_J$ with 
the inter-cluster hopping $\hat{\mcal{T}}(\vec{\mb{R}}_I - \vec{\mb{R}}_J)$. 
Hereafter, the cluster Hamiltonians are assumed to be identical, i.e., 
$\hat H_{\rm c}(\vec{\mb{R}}_I)=\hat H_{\rm c}$.

In the CPT, the exact single-particle Green's function matrix $\bs{G}'(z)$ of $H_{\rm c}$ 
on a cluster is numerically calculated. 
Therefore, the short-range correlations are taken into account exactly in $\bs{G}'(z)$, but the longer-range 
correlations beyond the size of the cluster should be approximated. 
Applying the strong coupling expansion 
with respect to the inter-cluster hopping $\hat{\mcal{T}}$, 
the single-particle Green's function matrix $\bs G\ofkz$ of the Hamiltonian $\hat{H}$ on the infinite lattice 
is obtained as 
\begin{equation}\label{eq.cpt}
  G_{\alpha \beta}\ofkz = \frac{1}{L_{\rm c}} \sum_{i=1}^{L_{\rm c}} \sum_{j=1}^{L_{\rm c}} 
  \tilde{G}_{i\alpha,j\beta}\ofkz \e^{-\imag \vec{\mb{k}} \cdot (\vec{\mb{r}}_i - \vec{\mb{r}}_j)}, 
\end{equation}
where $\vec{\mb{r}}_i$ is the position of site $i\,(=1,2,\cdots,L_{\rm c})$ inside the cluster 
($L_{\rm c}$: the number of sites in the single cluster), and 
the spin and orbital are labeled by $\alpha\,(=1,\cdots,O_{\rm c})$. 
Notice here that a site in the infinite lattice is specified with the cluster to which the site belongs and the 
location of the site within the cluster, i.e., site $i$ in the $I$-th cluster being represented as 
\begin{equation}
\vec{\mb X}_{i,I}=\vec{\mb{r}}_i+\vec{\mb{R}}_I.
\label{eq:position}
\end{equation}
$\tilde{G}_{i\alpha,j\beta}\ofkz$ is evaluated from the single-particle Green's function matrix $\bs{G}'(z)$ of 
the cluster: 
\begin{equation}\label{eq.Gcpt}
  \tilde{\bs{G}}\ofkz= \left[{\bs{G}'\ofz}^{-1} - \bs{\mcal{T}}\ofk \right]^{-1}.  
\end{equation}
Here, $\bs{\mcal{T}}\ofk$ is a matrix representation of $\hat{\mcal{T}}$ in the momentum $\vec{\mb{k}}$ space 
and is given as 
\begin{equation}
  \label{Vij}
  \mcal{T}_{i\alpha, j\beta}\ofk = \sum_{J\,(\ne I)} t_{i\alpha,j\beta}^{I,J}  \e^{\imag \vec{\mb{k}} \cdot ( \vec{\mb{R}}_I- \vec{\mb{R}}_J) },
\end{equation}
where $t_{i\alpha,j\beta}^{I,J}$ is the hopping integral between site $i$ in the $I$-th cluster 
located at $ \vec{\mb{R}}_I$ 
with spin-orbital $\alpha$ 
and site $j$ in the $J$-th cluster 
located at $ \vec{\mb{R}}_J$ 
with spin-orbital $\beta$, and the sum over $J$ excludes $I$.

The single-particle Green's function $\bs{G}'(z)$ of the cluster 
at temperature $T$ is  
\begin{equation}
  G'_{i\alpha, j\beta} \ofz = \sum_{m=1}^{N_{\rm pole}}
  \frac{Q_{i\alpha, m} Q_{j \beta, m}^*}{z - \lambda_m}, 
  \label{eq:gp}
\end{equation}
where 
\begin{equation}
  Q_{i\alpha,m} = 
  \sqrt{\frac{\e^{-E_r/T} + \e^{-E_s/T}}{Z}} 
  \bra r |\hat{c}_{\vec{\mb{r}}_i\alpha} | s \ket  
  \label{eq:q}
\end{equation}
and 
\begin{equation}
Z=\sum_r\e^{-E_r/T}
\label{eq:z}
\end{equation}
is the partition function of the cluster. 
In the above equations, $\hat{c}_{\vec{\mb{r}}_i\alpha}$ is the annihilation operator of fermion at site $i$ 
with spin-orbital $\alpha$, $| r \ket$ and $|s\ket$ are the many-body eigenstates of the cluster Hamiltonian 
$\hat{H}_{\rm c}$ with the eigenvalues $E_r$ and $E_r$, respectively, 
$\lambda_m = E_r - E_s$, 
and $m=(r,s)=1,2, \cdots, N_{\rm pole}$ labels all possible single-particle excitations~\cite{Aichhorn2006}. 
The many-body eigenstates and eigenvalues of the cluster Hamiltonian $\hat{H}_{\rm c}$ as well as 
the partition function $Z$ of the cluster are calculated using, e.g., the numerically exact diagonalization method. 

It is apparent in Eqs.~(\ref{eq:gp})--(\ref{eq:z}) that the temperature dependence is carried 
solely through $Q_{i \alpha,m}$. 
Therefore, the CPT at zero temperature is obtained simply  
by setting $Q_{i \alpha,m}$ in Eq.~(\ref{eq:q}) in the zero temperature limit~\cite{Aichhorn2006}, i.e.,   
\begin{equation}
  Q_{i\alpha,m} = 
  \left(\delta_{r0} + \delta_{s0} \right) \bra r | \hat{c}_{\vec{\mb{r}}_{i} \alpha} |s \ket,  
\end{equation}
where ``$0$" in Kronecker deltas $\delta_{r0}$ and $\delta_{s0}$ represents the ground state $| 0 \ket$ 
of the cluster Hamiltonian $\hat{H}_{\rm c}$. 
We have assumed here that the ground state $| 0 \ket$ is 
not degenerate. However, the extension to the case where the ground state of the cluster Hamiltonian 
$\hat{H}_{\rm c}$ is degenerate is straightforward. 
Notice that the order of $N_{\rm pole}$ at zero temperature is approximately the square root of that 
at finite temperatures if no truncation approximation for higher energy excitations is employed. 

\section{Formalism}\label{sec:formalism}

\subsection{Lehmann representation} \label{sec:lehmann}

Following Refs.~\cite{Senechal2012,Zacher2002,Aichhorn2006}, 
we derive the Lehmann representation of  
the single-particle Green's function $\bs G\ofkz$ of $\hat H$ on the infinite lattice given in Eq.~(\ref{eq.cpt}). 
This formulation allows us to calculate the spectral-weight functions 
and single-particle excitation energies explicitly. 

In the matrix notation, the single-particle Green's function $\bs{G}'(z)$ of the cluster can be written as 
\begin{equation}\label{eq.GcLehmann}
  \bs{G}'\ofz = \bs{Q}\left(z - \bs{\Lambda} \right)^{-1} \bs{Q}^\dag, 
\end{equation}
where $\bs{Q}$ is the $L_{\rm c}O_{\rm c} \times N_{\rm pole}$ matrix with the matrix elements 
$Q_{i\alpha,m}$ defined in Eq.~(\ref{eq:q})  
and $\bs{\Lambda} = {\rm diag}(\lambda_1, \lambda_2,\cdots, \lambda_{N_{\rm pole}})$. 
Substituting Eq.~(\ref{eq.GcLehmann}) into Eq.~(\ref{eq.Gcpt}) yields 
\begin{equation}
  \tilde{\bs{G}}\ofkz = \bs{Q} \left[z - \bs{M}\ofk \right]^{-1} \bs{Q}^\dag,  
\end{equation}
where we have introduced an $N_{\rm pole} \times N_{\rm pole}$ Hermitian matrix
\begin{equation} \label{eq.M}
  \bs{M}\ofk = \bs{\Lambda} + \bs{Q}^\dag \bs{\mcal{T}}\ofk \bs{Q}.
\end{equation}
Since $\bs{M}\ofk$ is Hermitian, this matrix is diagonalized by a unitary matrix $\bs{U}\ofk$ as 
\begin{eqnarray}\label{eq.diagM}
  \Tilde{\bs{\Lambda}}\ofk 
  &=&   \bs{U}^\dag \ofk \bs{M}\ofk \bs{U}\ofk \notag \\
  &=& {\rm diag}{\left[\w_1\ofk,\w_2\ofk,\cdots, \w_{N_{\rm pole}} \ofk \right]}. 
\end{eqnarray} 
Therefore, $\tilde{\bs{G}}(\vec{\mb{k}},z)$ in Eq.~(\ref{eq.Gcpt}) is given as  
\begin{equation}
  \tilde{\bs{G}}(\vec{\mb{k}},z) 
  = \Tilde{\bs{Q}}\ofk \left[z - \Tilde{\bs{\Lambda}}\ofk \right]^{-1} \Tilde{ \bs{Q}}^\dag\ofk,  
\end{equation} 
where $\Tilde{\bs{Q}}\ofk = \bs{Q} \bs{U}\ofk$.

The Lehmann representation of 
the translationally invariant single-particle Green's function $\bs G\ofkz$ of $\hat H$ 
on the infinite lattice in Eq.~(\ref{eq.cpt}) is thus obtained as 
\begin{equation}\label{eq:Gab}
  G_{\alpha \beta}\ofkz = \sum_{m = 1}^{N_{\rm pole}} \frac{ A_{\alpha \beta, m}\ofk }{z - \w_m\ofk}, 
\end{equation}
where the spectral-weight function $A_{\alpha \beta, m} \ofk$ is given as
\begin{equation}\label{eq.Aab}
  A_{\alpha \beta, m} \ofk = \frac{1}{L_{\rm c}} \sum_{i=1}^{L_{\rm c}} \sum_{j=1}^{L_{\rm c}}   
  \tilde{Q}_{i \alpha, m}\ofk \tilde{Q}_{j \beta, m}^*\ofk \e^{-\imag \vec{\mb{k}} \cdot (\vec{\mb{r}}_i  -  \vec{\mb{r}}_j)}, 
\end{equation} 
and the single-particle excitation energies $\w_m \ofk$  
correspond to the eigenvalues of $\bs{M}\ofk$ in Eq.~(\ref{eq.diagM}).  
Note that $A_{\alpha \beta, m}\ofk$ fulfills 
the spectral-weight sum rule for each momentum $\vec{\mb k}$~\cite{Senechal2008}, i.e.,  
\begin{equation}\label{eq.sumrule}
  \sum_{m=1}^{N_{\rm pole}} A_{\alpha \beta, m} \ofk = \delta_{\alpha \beta}.   
\end{equation}

Once the single-particle Green's function $\bs G\ofkz$ is obtained, 
the single-particle excitation spectrum $A_{\alpha,\beta}(\vec{\mb{k}},\w)$ of $\hat H$ on the infinite lattice 
is readily calculated as 
\begin{equation}
A_{\alpha\beta}(\vec{\mb{k}},\w) = -\frac{1}{\pi}\lim_{\eta\to0} {\rm Im} G_{\alpha \beta}(\vec{\mb{k}},\w+\imag\eta).
\end{equation} 
and the density of states $\rho_{\alpha\beta}(\w)$ is simply obtained 
by integrating $A_{\alpha\beta}(\vec{\mb{k}},\w)$ over the momentum $\vec{\mb{k}}$ in the whole  Brillouin zone. 
In the practical numerical calculations, we usually take a finite but small value of $\eta$, which 
corresponds to the Lorentzian broadening factor of delta-function peaks. 
The frequency-dependent broadening factor $\eta(\w)$ was also introduced to better control the 
high energy structures of the spectrum~\cite{Civelli2009}.
However, these procedures are not suitable  
when we examine fine structures of the spectrum 
because they are obscured by the tails of Lorentzian functions.

\subsection{Effective single-particle Hamiltonian for single-particle excitations} 

\begin{figure}
  \begin{center}
    \includegraphics[width=7.7cm]{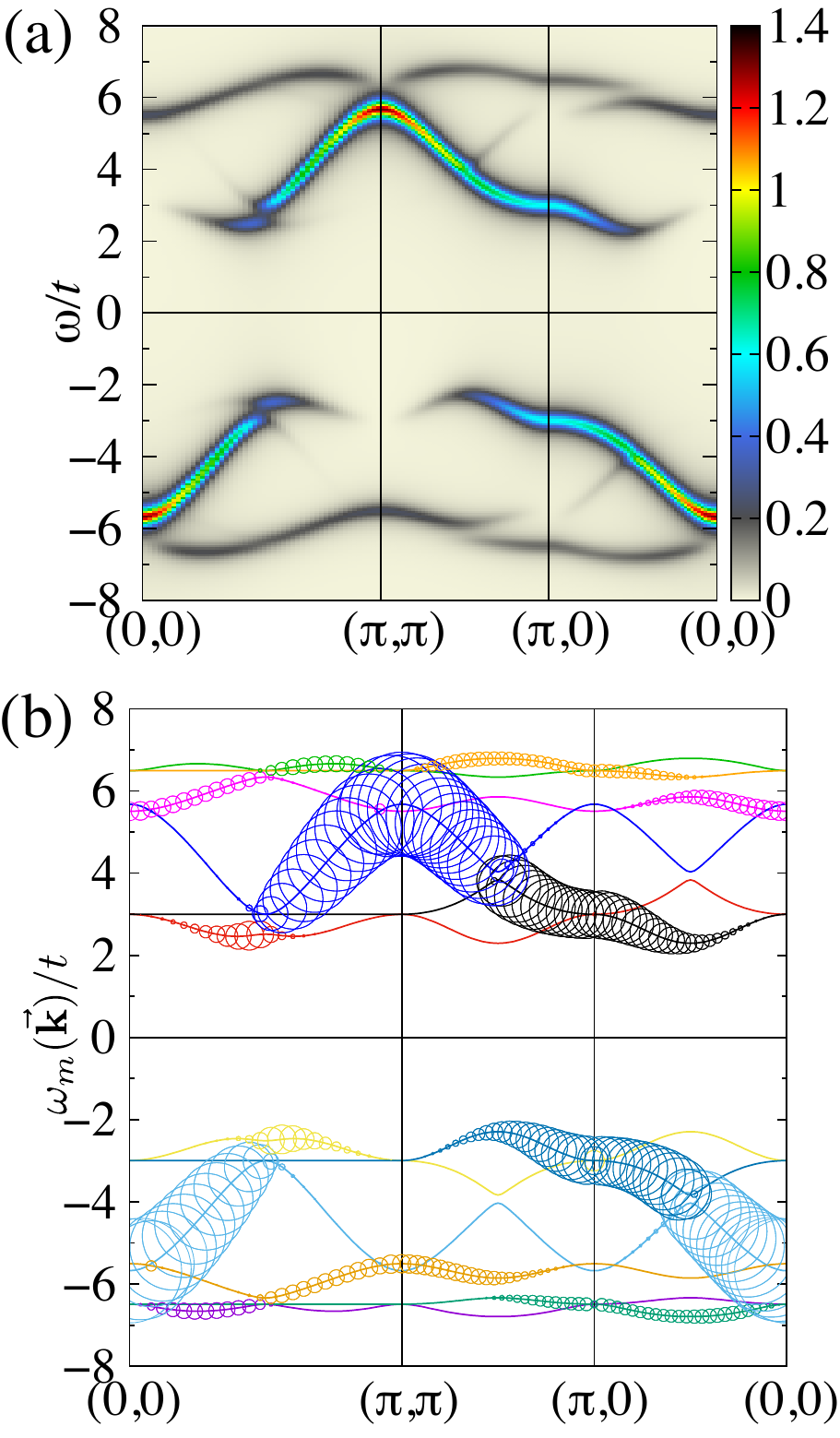}
    \caption{
      \label{fig.bands}
      (a) The single-particle excitation spectrum $-{\rm Im} G_{\alpha \alpha}(\vec{\mb{k}},\w+\imag\eta)/\pi$ 
      with the Lorentzian broadening of $\eta/t=0.2$ and 
      (b) the ``band structure'' of the corresponding effective single-particle Hamiltonian 
      for the single-band Hubbard model 
      defined in Eq.~(\ref{eq:HB}) on the square lattice at half-filling and $T=0$ with 
      the on-site Coulomb repulsion $U/t=8$. 
      The results are obtained by the CPT using a $2 \times 2$ cluster. 
      Although the total number $N_{\rm pole}$ of ``energy bands" $\w_m(\vec{\mb{k}})$ 
      is $48$ for this cluster 
      (considering only the conservation of the particle number and the $z$-component 
      of the total spin), 
      only the energy bands with 
      non-zero spectral weight are shown in (b), where 
      the radius of each circle is proportional to the spectral-weight function 
      $A_{\alpha \alpha,m}\ofk$.  
      Different colors in (b) indicate different bands, 
      where the band connectivity is properly resolved (see Sec.~\ref{sec:bc}). 
    }
  \end{center}
\end{figure}

It is now clear from the Lehmann representation in Eq.~(\ref{eq:Gab}) that 
apart from the spectral weight 
the structure of the single-particle Green's function $\bs G\ofkz$ of interacting fermions 
is identical with that of 
a non-interacting many $O_{\rm c} N_{\rm pole}$-orbital system. 
In this sense, the Hermitian matrix $\bs{M}(\vec{\mb{k}})$ in Eq.~(\ref{eq.M}) 
can be regarded as an effective single-particle Hamiltonian 
since the eigenvalues of $\bs{M}(\vec{\mb{k}})$ coincide with the single-particle excitation energies 
for the interacting fermions. 

Figure~\ref{fig.bands} shows an example of the comparison between the single-particle excitation 
spectrum of the single-band Hubbard model  
and the eigenvalues of the corresponding effective single-particle Hamiltonian $\bs{M}(\vec{\mb{k}})$. 
It is apparent in Fig.~\ref{fig.bands} that the single-particle excitation energies 
of interacting fermions are indeed given as the eigenvalues of the corresponding effective 
single-particle Hamiltonian. 
In appendix~\ref{appendix}, we 
establish the general correspondence between 
the self-energy and the matrix elements of the effective single-particle Hamiltonian.

It should be noted here that the notion of an effective single-particle Hamiltonian for the single-particle excitations 
in the interacting single-particle Green's function within the CPT has already 
been noticed in appendix of Ref.~\cite{Zacher2002}.  
A similar notion has also been introduced recently in 
the cofermion or hidden-fermion description 
of the single-particle excitations 
to construct an approximate single-particle Hamiltonian for the single-particle excitations of 
the 2D Hubbard models~\cite{Yamaji2011PRL,Yamaji2011PRB,Imada2011,Sakai2015,Sakai2014}.

Since there exists the correspondence between the single-particle excitation energies of interacting 
fermions and the effective single-particle Hamiltonian, 
we can now employ the tetrahedron method, which has been developed 
in the single-particle theory~\cite{Rath1975,Blochl1994}, 
to evaluate the density of states of interacting fermions. 
Namely, applying the tetrahedron method, we can perform with high accuracy the momentum integral 
of the interacting single-particle excitation spectrum $A_{\alpha\beta}(\vec{\mb k},\w)$ over the whole Brillouin zone.

\subsection{Tetrahedron method}

The tetrahedron method for the single-particle theory in 
3D systems has been described in details in Refs.~\cite{Rath1975,Blochl1994}. 
Here, we describe the tetrahedron method for 2D systems since 
many interesting classes of interacting fermion systems are included, such as 
interacting Dirac fermions~\cite{Seki,Hassan2013,Yamada2015,Ebato2015,Meng2010,Sorella2012,Assaad2013,Toldin2015,Otsuka2016}, 
high-$T_{\rm c}$ cuprate superconductors 
which exhibit superconductivity as well as pseudogap phenomena 
in the lightly hole doped regime~\cite{Senechal2004,
Tohyama2004,Kyung2006,Aichhorn2007,Kancharla2008,Sakai2009,Eder2010,Eder2011,Kohno2012,Sakai2013,Kohno2014,Sakai2014}, 
and various interface electron states in strongly correlated heterostructures~\cite{Potthoff1999,Liebsch2003,Yunoki2007,Charlebois2013}.

\subsubsection{Triangular partitioning of Brillouin zone }\label{sec:part_BZ}

We consider a 2D Brillouin zone defined by 
the reciprocal lattice vectors $\vec{\mb{G}}_1$ and $\vec{\mb{G}}_2$, 
and divide the Brillouin zone into $N_1  N_2$ parallelograms.  
Each parallelogram is defined by the two vectors 
\begin{eqnarray}
  \vec{\mb{g}}_1 = \vec{\mb{G}}_1/N_1
\end{eqnarray}
and 
\begin{eqnarray}
  \vec{\mb{g}}_2 = \vec{\mb{G}}_2/N_2.  
\end{eqnarray}
We further divide each parallelogram into two triangles. 
The total number  $N_T$ of triangles is thus 
\begin{equation}
  N_T = 2 N_1 N_2.  
\end{equation}
The volume $V_G$ of the Brillouin zone and the volume $V_T$ of the triangle are given as 
\begin{eqnarray}
  V_{G} = \left|\det{\left(\vec{\mb{G}}_1, \vec{\mb{G}}_2\right)} \right| 
\end{eqnarray}
and 
\begin{eqnarray}
  V_T = \frac{1}{2} \left|\det{\left(\vec{\mb{g}}_1, \vec{\mb{g}}_2 \right)} \right| = \frac{V_{G}}{N_T},    
\end{eqnarray} 
respectively.

\subsubsection{Density of states}

Using the single-particle Green's function $\bs G\ofkz$ on the infinite lattice in Eq.~(\ref{eq:Gab}), 
the density of states $\rho_{\alpha \beta}(\w)$ per unit cell 
projected onto spin-orbitals $\alpha$ and $\beta$ is obtained as 
\begin{eqnarray}
  \rho_{\alpha \beta}(\w) 
  &=&\lim_{\eta \to 0} \frac{1}{V_{G}}\int_{\rm BZ} \dd^2 k \left[-\frac{1}{\pi}{\rm Im} G_{\alpha \beta}(\vec{\mb{k}},\w+\imag \eta) \right] \notag \\
  &=&\lim_{\eta \to 0} \sum_{m=1}^{N_{\rm pole}} \frac{1}{V_{G}}\int_{\rm BZ} \dd^2 k 
  \left[\frac{1}{\pi}\frac{A_{\alpha \beta,m}\ofk \ \eta}{\left(\w - \w_m(\vec{\mb{k}})\right)^2+\eta^2} \right]\notag \\
  &=&\sum_{m=1}^{N_{\rm pole}} \frac{1}{V_{G}}\int_{\rm BZ} \dd^2 k A_{\alpha \beta,m}(\vec{\mb{k}}) 
  \delta\left(\w - \w_m(\vec{\mb{k}})\right), 
  \label{eq:dos}
\end{eqnarray}
where the momentum $\vec{\mb k}$ integral is performed over the whole 2D Brillouin zone. 
In the third equality, we have used the fact that the Lorentzian function in the limit of $\eta \to 0$ 
becomes the delta function. 
In appendix~\ref{appendixB}, we show that in the CPT the 
density of states can also be calculated with the integral over the 
reduced Brillouin zone for the superlattice on which the clusters are defined.

As in the tetrahedron method~\cite{Rath1975,Blochl1994}, 
we first recast the integral over the 2D Brillouin zone in Eq.~(\ref{eq:dos})
to the sum of integrals over small triangles $\triangle_\tau$ 
with $\tau=1,2,\dots,N_T$ covering the 2D Brillouin zone, i.e., 
\begin{equation}
  \frac{1}{V_{G}} \int_{\rm BZ} \dd^2 k \cdots = \sum_{\tau=1}^{N_{T}} \frac{1}{V_{G}} \int_{\triangle_\tau} \dd^2 k \cdots.   
  \label{eq:int_tetra}
\end{equation}
To perform the $\vec{\mb{k}}$ integral analytically over each small triangle, 
the single-particle excitation energies $\w_{m}\ofk$ 
and the spectral-weight functions $A_{\alpha\beta,m}\ofk$ are regarded 
as a linear function of momentum $\vec{\mb{k}}$ within each triangle. 
The density of states is then expressed as 
\begin{equation} 
  \rho_{\alpha \beta}(\w) = 
  \sum_{m=1}^{N_{\rm pole}}  
  \sum_{\tau=1}^{N_T} 
  A_{\alpha \beta,  m, \tau}(\w)
  D_{m,\tau}(\w),  
  \label{eq:tetdos}
\end{equation}
where $D_{m,\tau}(\w)$ and $A_{\alpha \beta, m, \tau}(\w) $ are respectively 
the density of states and the spectral weight of spin-orbitals $\alpha$ and $\beta$ 
contributed from the $m$-th pole at triangle $\tau$. 
In the following, we derive the analytical expressions of $D_{m, \tau}(\w)$ and $A_{\alpha \beta, m, \tau}(\w)$.

\begin{figure}
  \begin{center}
    \includegraphics[width=7.8cm]{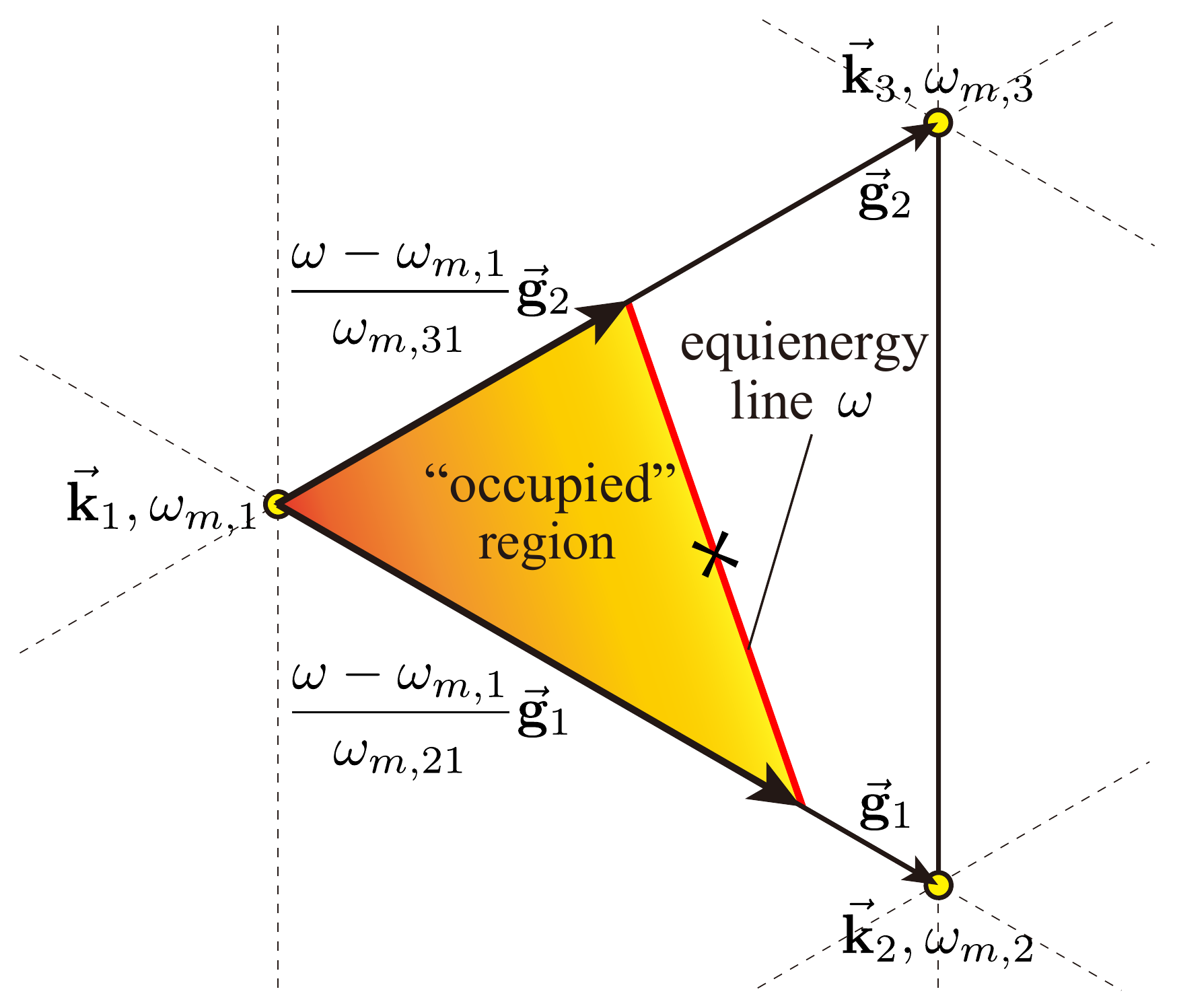}
    \caption{
      \label{fig.tetra}
      Illustration of the triangular partitioning (dashed lines) of the Brillouin zone. 
      A single triangle is highlighted by thick solid lines for the derivation of $n_{m,\tau}(\w)$, 
      assuming here that $\w_{m,1} \leqslant \w < \w_{m,2} \leqslant \w_{m,3}$ [see also Eq.~(\ref{eq.nw})]. 
      For a given energy $\w$, the equienergy line of $\w$ (red solid line) is determined uniquely 
      inside the triangle. 
      The cross on the equienergy line indicates the midpoint of the line. 
      The volume of the ``occupied'' region (shaded area) divided by $V_{G}$ is simply give as 
      $n_{m,\tau}(\w)=\frac{1}{2 V_{G}} \frac{(\w-\w_{m,1})^2}{\w_{m,21}\w_{m,31}} |\det{\left(\vec{\mb{g}}_1,\vec{\mb{g}}_2\right)}|$. 
    }
  \end{center}
\end{figure}

In order to derive $D_{m,\tau}(\w)$, 
we first consider the number of states $n_{m,\tau}(\w)$ per unit cell ``occupied" 
in the $m$-th band (i.e., pole dispersion) below $\w$ at the $\tau$-th triangle of the Brillouin zone.  
Let us define three momenta  
$\vec{\mb{k}}_{1}$, $\vec{\mb{k}}_{2}$, and $\vec{\mb{k}}_{3}$ 
on the corners of the $\tau$-th triangle.  
Here, we assume without loss of generality that the single-particle excitation energies are 
in ascending order at these momenta, i.e., 
$\w_{m}(\vec{\mb{k}}_{1}) \leqslant \w_{m}(\vec{\mb{k}}_{2}) \leqslant \w_{m}(\vec{\mb{k}}_{3})$.   
As depicted in Fig.~\ref{fig.tetra}, $n_{m,\tau}(\w) V_{G}$ is the volume of the ``occupied'' region 
in the $\tau$-th triangle. It is now easy to show that 
\begin{eqnarray}
  \label{eq.nw}
  n_{m, \rm{\tau}} (\w) =
  \left\{
    \begin{array}{ll}
      0 & (\w < \w_{m,1}) \\
      \displaystyle 
      \frac{\left(\w-\w_{m,1}\right)^2}{\w_{m,31}\w_{m,21}}\frac{V_T}{V_{G}} 
      & (\w_{m,1} \leqslant \w < \w_{m,2}) \\
      \displaystyle  
      \left[1-\frac{\left(\w_{m,3}-\w\right)^2}{\w_{m,31}\w_{m,32}}\right]\frac{V_T}{V_{G}}
      & (\w_{m,2} \leqslant \w < \w_{m,3}) \\
      \displaystyle \frac{V_T}{V_{G}}
      & (\w_{m,3} \leqslant \w)      \\
    \end{array},
    \right.
\end{eqnarray}
where $\w_{m,i} \equiv \w_{m}(\vec{\mb{k}}_{i})$ and $\w_{m,ij} = \w_{m,i} - \w_{m,j}$.  
Since $D_{m, \tau}(\w) = \partial n_{m,\tau} (\w)/\partial \w $, we find that 
\begin{eqnarray} 
  D_{m, \tau} (\w) = 
  \left\{
    \begin{array}{ll}
      0 & (\w < \w_{m,1}) \\
      \displaystyle 
      \frac{2\left(\w-\w_{m,1}\right)}{\w_{m,31}\w_{m,21}}\frac{V_T}{V_{G}} & (\w_{m,1} \leqslant \w < \w_{m,2}) \\
      \displaystyle  
      \frac{2\left(\w_{m,3}-\w\right)}{\w_{m,31}\w_{m,32}}\frac{V_T}{V_{G}} & (\w_{m,2} \leqslant \w < \w_{m,3}) \\
      0 & (\w_{m,3} \leqslant \w)      \\
    \end{array}.
    \right. 
\end{eqnarray}

The spectral weight $A_{\alpha\beta,m,\tau} (\w)$ in Eq.~(\ref{eq:tetdos}) is 
a weighted average of the spectral-weight function 
$A_{\alpha \beta,m} (\vec{\mb{k}_i})$ within the $\tau$-th triangle, i.e., 
\begin{equation}
  A_{\alpha \beta, m, \tau}(\w) = 
  \sum_{i=1}^{3} f_{m,\tau,i}(\w) A_{\alpha \beta, m} (\vec{\mb{k}}_i),  
\end{equation} 
where the averaging weight $f_{m,\tau,i}(\w)$ is determined 
in such as way that $A_{\alpha \beta, m, \tau}(\w)$ is 
the spectral-weight function averaged over 
the equienergy line, or equivalently, 
the spectral-weight function at the midpoint of the 
equienergy line, indicated by the cross in Fig.~{\ref{fig.tetra}}. 
The averaging weight $f_{m,\tau,i}(\w)$ is thus obtained as 
\begin{eqnarray}
  2f_{m,\tau,1} (\w) =
  \left\{
    \begin{array}{ll}
      0 & (\w < \w_{m,1}) \\
      \displaystyle \frac{\w_{m,2}-\w}{\w_{m,21}} + \frac{\w_{m,3}-\w}{\w_{m,31}} 
      & (\w_{m,1} \leqslant \w < \w_{m,2}) \\      
      \displaystyle \frac{\w_{m,3}-\w}{\w_{m,31}} 
      & (\w_{m,2} \leqslant \w < \w_{m,3}) \\
      0 & (\w_{m,3} \leqslant \w)      \\
    \end{array},
  \right. 
\end{eqnarray}
\begin{eqnarray}
  2f_{m,\tau,2} (\w) =
  \left\{
    \begin{array}{ll}
      0 & (\w < \w_{m,1}) \\
      \displaystyle \frac{\w-\w_{m,1}}{\w_{m,21}} 
      & (\w_{m,1} \leqslant \w < \w_{m,2}) \\
      \displaystyle \frac{\w_{m,3}-\w}{\w_{m,32}} 
      & (\w_{m,2} \leqslant \w < \w_{m,3}) \\
      0 & (\w_{m,3} \leqslant \w)      \\
    \end{array},
    \right. 
\end{eqnarray}
and 
\begin{eqnarray}
  2f_{m,\tau,3} (\w) =
  \left\{
    \begin{array}{ll}
      0 & (\w < \w_{m,1}) \\
      \displaystyle \frac{\w - \w_{m,1}}{\w_{m,31}}
      & (\w_{m,1} \leqslant \w < \w_{m,2}) \\
      \displaystyle \frac{\w - \w_{m,1}}{\w_{m,31}} + \frac{\w-\w_{m,2}}{\w_{m,32}}
      & (\w_{m,2} \leqslant \w < \w_{m,3}) \\
      0 & (\w_{m,3} \leqslant \w)      \\
    \end{array}.
    \right. 
\end{eqnarray}
Note that the averaging weight $f_{m,\tau,i}(\w)$ fulfills that 
\begin{equation}
\sum_{i=1}^3 f_{m,\tau,i}(\w) =
\left\{
\begin{array}{ll}
0 & (\w < \w_{m,1}) \\
1 & (\w_{m,1} \leqslant \w < \w_{m,3}) \\
0 & (\w_{m,3} \leqslant \w)
\end{array}. 
\right.
\end{equation}
Table~\ref{table} summarizes the functions 
necessary to the tetrahedron method for 2D systems.

\begin{table*}
  \caption{
    \label{table}
    Functions used in the tetrahedron method combined with the CPT for 2D systems. 
    $D_{m,\tau}(\w)$ and $f_{m,\tau,i}(\w)$ are required for the calculation of density of states, and 
    $\w_{m,\tau,i}(\w)$ is for the grand potential functional in the VCA.   
  }
  \begin{tabular}{ccccc} 
    \hline
    \hline
    Functions & $\w < \w_{m,1}$ & $\w_{m,1} \leqslant \w < \w_{m,2}$ & $\w_{m,2} \leqslant \w < \w_{m,3}$ & $\w_{m,3} \leqslant \w$ \\
      \hline 
    $n_{m,\tau}(\w)$  & 
    0  & 
    $\displaystyle \frac{\left(\w-\w_{m,1}\right)^2}{\w_{m,31}\w_{m,21}}\frac{V_T}{V_{G}}$ & 
    $\displaystyle \left[1-\frac{\left(\w_{m,3}-\w\right)^2}{\w_{m,31}\w_{m,32}}\right]\frac{V_T}{V_{G}}$   & 
    $\displaystyle \frac{V_T}{V_G}$ \\ 
    \\
    $D_{m,\tau}(\w)$ & 
    0  & 
    $\displaystyle \frac{2\left(\w-\w_{m,1}\right)}{\w_{m,31}\w_{m,21}}\frac{V_T}{V_{G}}$ & 
    $\displaystyle \frac{2\left(\w_{m,3}-\w\right)}{\w_{m,31}\w_{m,32}}\frac{V_T}{V_{G}}$  & 
    0 \\
    \\
    $f_{m,\tau,1}(\w)$ & 
    0 & 
    $\displaystyle \frac{1}{2}\left(\frac{\w_{m,2}-\w}{\w_{m,21}} + \frac{\w_{m,3}-\w}{\w_{m,31}} \right)$ & 
    $\displaystyle \frac{1}{2}\frac{\w_{m,3}-\w}{\w_{m,31}} $ & 
    0 \\
    $f_{m,\tau,2}(\w)$ & 
    0 & 
    $\displaystyle \frac{1}{2} \frac{\w-\w_{m,1}}{\w_{m,21}}$ &
    $\displaystyle \frac{1}{2} \frac{\w_{m,3}-\w}{\w_{m,32}}$ &
    0 \\
    $f_{m,\tau,3}(\w)$ & 
    0 &
    $\displaystyle \frac{1}{2} \frac{\w - \w_{m,1}}{\w_{m,31}}$ &
    $\displaystyle \frac{1}{2} \left(\frac{\w - \w_{m,1}}{\w_{m,31}} + \frac{\w-\w_{m,2}}{\w_{m,32}}\right)$ &
    0 \\
    \\
    $w_{m,\tau,1}(\w)$ & 
    0 & 
    $\displaystyle \frac{n_{m,\tau}(\w)}{3} \left(1 + \frac{\w_{m,2}-\w}{\w_{m,21}} + \frac{\w_{m,3}-\w}{\w_{m,31}} \right)$ & 
    $\displaystyle \frac{n_{m,\tau}(\w)}{3} \frac{\w_{m,3} - \w}{\w_{m,31}} $ &
    $\displaystyle \frac{n_{m,\tau}(\w)}{3}$\\
    $w_{m,\tau,2}(\w)$ & 
    0 & 
    $\displaystyle \frac{n_{m,\tau}(\w)}{3} \frac{\w - \w_{m,2}}{\w_{m,21}} $ &
    $\displaystyle \frac{n_{m,\tau}(\w)}{3} \frac{\w_{m,3} - \w}{\w_{m,32}} $ & 
    $\displaystyle \frac{n_{m,\tau}(\w)}{3}$\\
    $w_{m,\tau,3}(\w)$ & 
    0 & 
    $\displaystyle \frac{n_{m,\tau}(\w)}{3} \frac{\w - \w_{m,3}}{\w_{m,31}} $ &
    $\displaystyle \frac{n_{m,\tau}(\w)}{3} \left(1 + \frac{\w-\w_{m,1}}{\w_{m,31}} + \frac{\w - \w_{m,2}}{\w_{m,32}} \right)$ & 
    $\displaystyle \frac{n_{m,\tau}(\w)}{3}$\\
    \hline
    \hline
  \end{tabular}
\end{table*}

Notice that the averaging weight $f_{m,\tau,i}(\w)$ 
is {\it not} the 2D analogue 
of the ``integration weight $w$'' in Ref.~\cite{Blochl1994}. 
This is because $f_{m,\tau,i}(\w)$ is introduced for the momentum integral of delta functions 
over the whole Brillouin zone, 
while the ``integration weight $w$'' is introduced for the momentum integral of step functions 
(i.e., the Fermi distribution function in the zero-temperature limit). 
The 2D analogue of ``$w$'' should be 
the integration weight for the average of a 
linearly approximated function $X_{m}\ofk$ over the occupied region. 
This results in the integration weight $w_{m,\tau,i}(\w)$ 
determined so as to give the value of $X_{m}\ofk$ at the center 
of the occupied region in the $\tau$-th triangle.   
Although the integration weight $w_{m,\tau,i}(\w)$ is not required for 
the calculation of density of states, we also show 
$w_{m,\tau,i}(\w)$ for 2D systems  
in Table~\ref{table} for further applications, which include the calculation of the grand potential functional 
(see Sec.~\ref{sec_vca}).

\subsubsection{Band connectivity}\label{sec:bc}  
As described in Ref.~\cite{Pickard1999}, 
the tetrahedron method sometimes induces 
artificial spikes or gaps in the density of states 
when the band crossing is not properly treated. 
To overcome this difficulty, we follow the prescription described 
in Ref.~\cite{Yazyev2002}   
and resolve the band connectivity. 

We first define the overlap matrix 
\begin{equation}
  \bs{F}(\vec{\mb{k}},\vec{\mb{k}}') = \bs{U}^\dag \ofk \bs{U}(\vec{\mb{k}'}),
\end{equation} 
where $\vec{\mb{k}}$ and $\vec{\mb{k}'}$ are two momenta close to each other. 
We assume that the eigenvectors of $\bs{M}\ofk$ are sorted in $\bs{U}\ofk$ 
according to the corresponding eigenenergies ordered ascendingly 
at each $\vec{\mb{k}}$ [see Eq.~(\ref{eq.diagM})]. 
When no band crossing occurs between the $m$-th and $n$-th bands, 
$|F_{mn}(\vec{\mb{k}},\vec{\mb{k}'})| \simeq \delta_{mn}$ because 
$\vec{\mb{k}}$ and $\vec{\mb{k}'}$ are close to each other, where 
$F_{mn}(\vec{\mb{k}},\vec{\mb{k}'})=\sum_{l=1}^{N_{\rm pole}}   \left[\bs{U}(\vec{\mb{k}})\right]_{lm}^* \left[\bs{U}(\vec{\mb{k}'})\right]_{ln}$. 
When a band crossing occurs between the $m$-th and $n$-th bands, 
$|F_{mn}(\vec{\mb{k}},\vec{\mb{k}'})| \simeq |F_{nm}(\vec{\mb{k}},\vec{\mb{k}'})| \simeq 1$ and 
$|F_{mm}(\vec{\mb{k}},\vec{\mb{k}'})| \simeq |F_{nn}(\vec{\mb{k}},\vec{\mb{k}'})| \simeq 0$. 
Therefore, we can systematically detect the band crossing from the overlap matrix elements 
$F_{mn}(\vec{\mb{k}},\vec{\mb{k}}')$. 

In the practical calculations, we can define that the band crossing occurs between the $m$-th 
and $n$-th bands when $|F_{mn}(\vec{\mb{k}},\vec{\mb{k}}')|^2 > 0.5$ for $m \not = n$. 
If the band crossing is detected, the excitation energies 
$\w_m\ofk$ and $\w_n(\vec{\mb{k}'})$ are assigned to the $m$-th band, and 
$\w_n\ofk$ and $\w_m(\vec{\mb{k}'})$ to the $n$-th band. 
An example of the connectivity resolved band structure is shown in Fig.~\ref{fig.bands}(b).  
After resolving the band connectivity, we can safely apply the tetrahedron method.

\section{Benchmark calculation} \label{sec:benchmark}

To demonstrate the method, we calculate the density of 
states of the single-band Hubbard model on the square lattice 
defined by the following Hamiltonian: 
\begin{equation}
  \hat{H} = 
  \sum_{\bra i,j \ket ,\s} t_{ij} 
  \left( \hat{c}_{i\s}^\dag \hat{c}_{j \s} + {\rm H. c.} \right) 
  + U \sum_{i} \hat{n}_{i \up} \hat{n}_{i \dn} 
  - \mu \sum_{i, \s} \hat{n}_{i \s},
  \label{eq:HB}
\end{equation}
where $\hat{c}_{i\s}^\dag$ creates an electron on site $i$ 
with spin $\s(=\up,\dn)$, 
and $\hat{n}_{i \s} = \hat{c}^\dag_{i \s} \hat{c}_{i \s}$. 
The sum indicated as $\bra i,j \ket $ runs over a pair of sites $i$ and $j$ with the hopping integral 
$t_{ij}$. 
Here, we set $t_{ij} = t$ when site $i$ is nearest 
neighbor to site $j$ and $t_{ij}=0$ otherwise. 
$U\,(>0)$ is the on-site Coulomb interaction and 
$\mu$ is the chemical potential.

\begin{figure}
  \begin{center}
    \includegraphics[width=6.75cm]{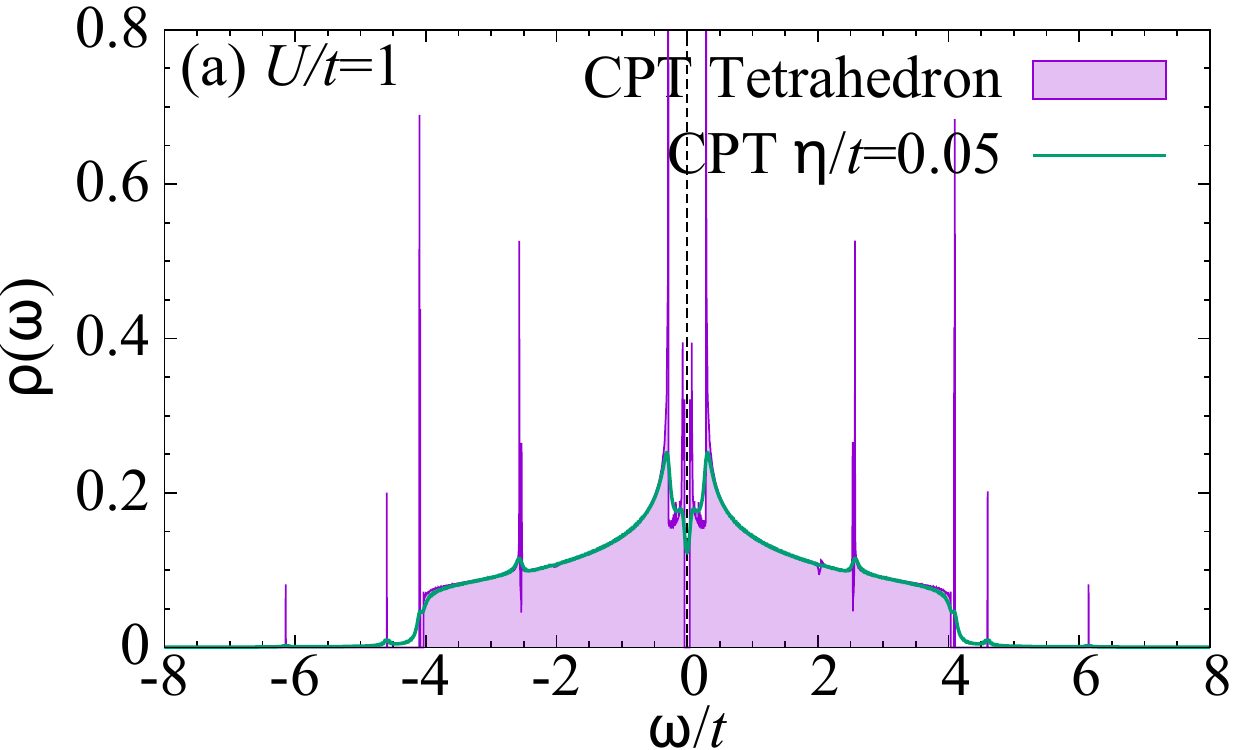}
    \includegraphics[width=6.75cm]{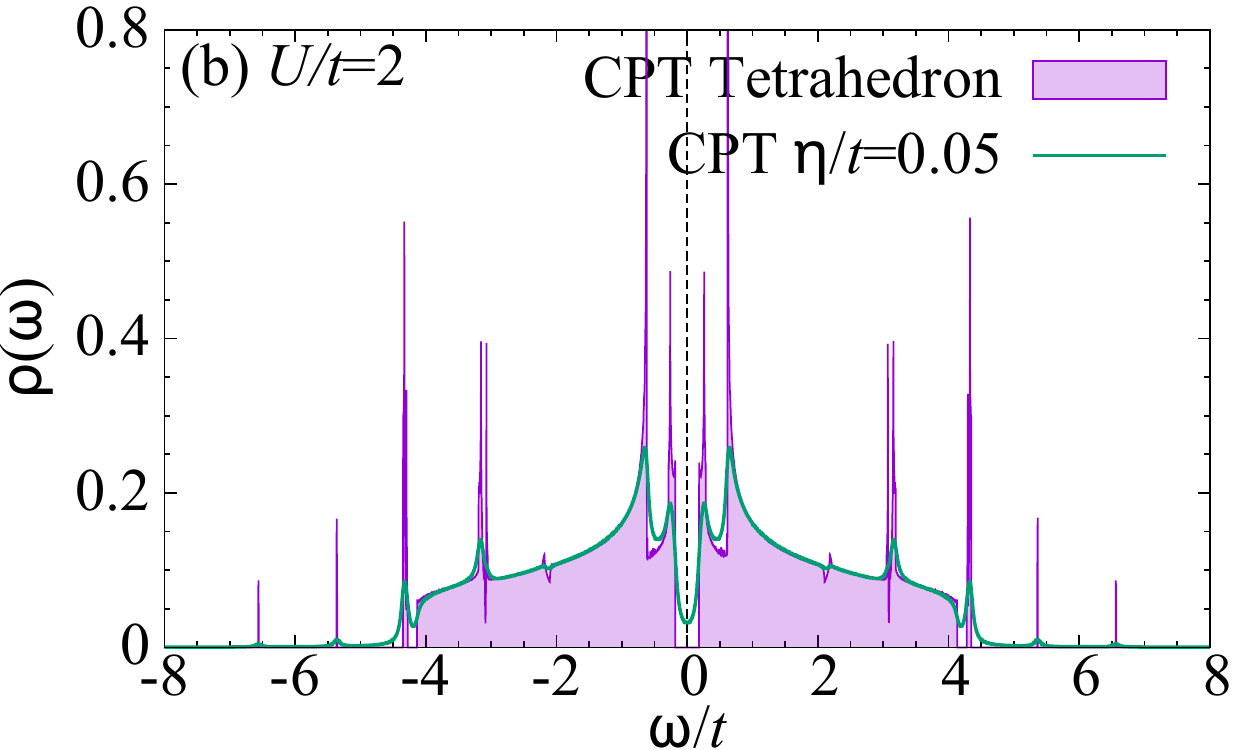}
    \includegraphics[width=6.75cm]{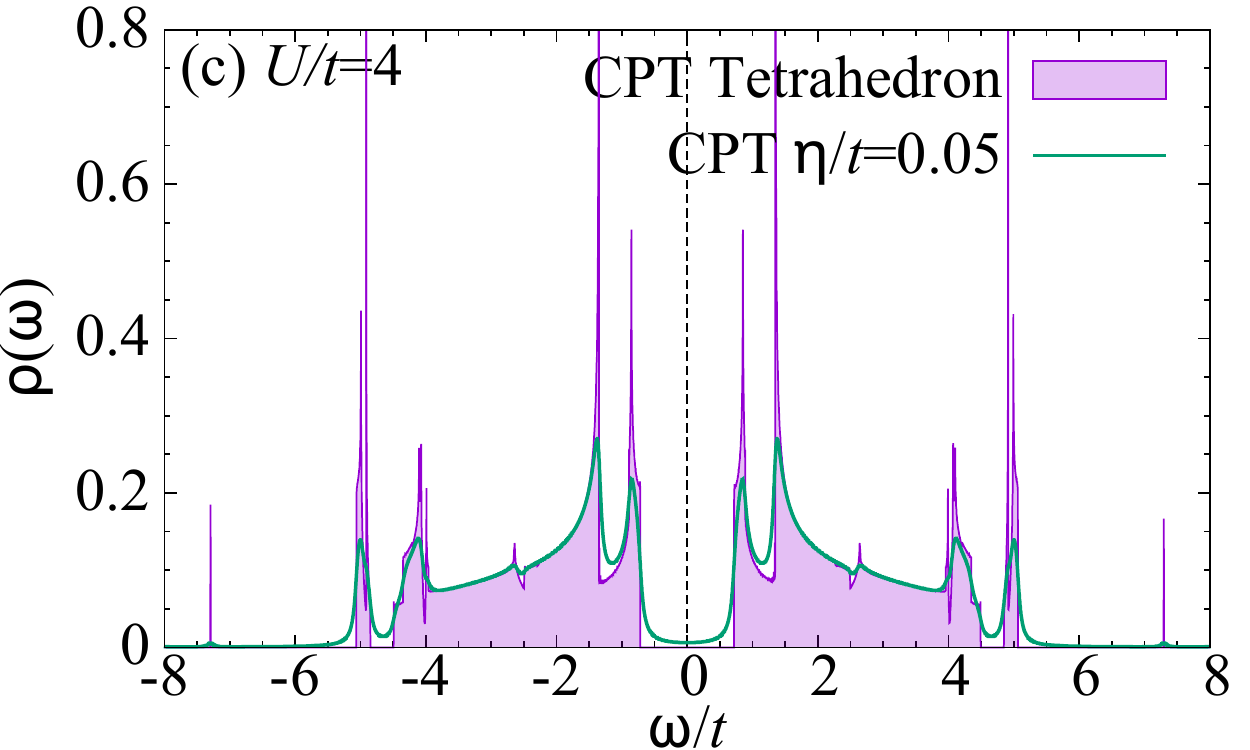}
    \includegraphics[width=6.75cm]{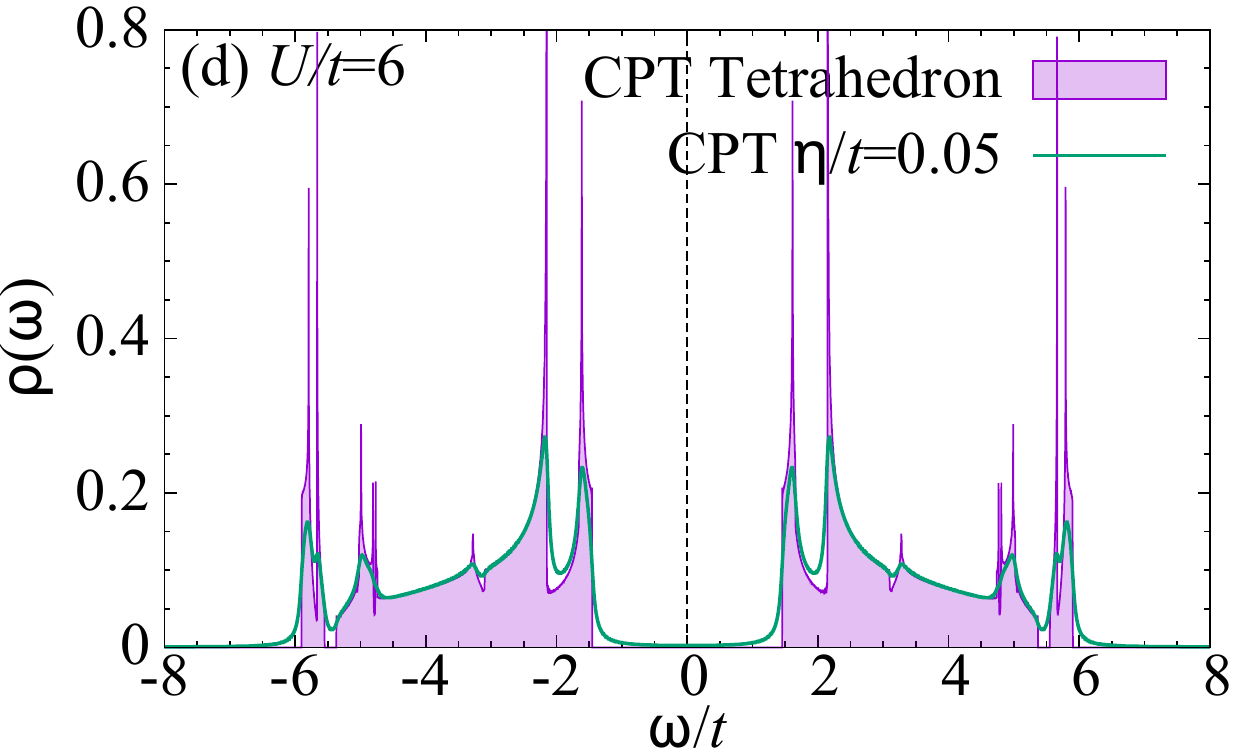}
    \includegraphics[width=6.75cm]{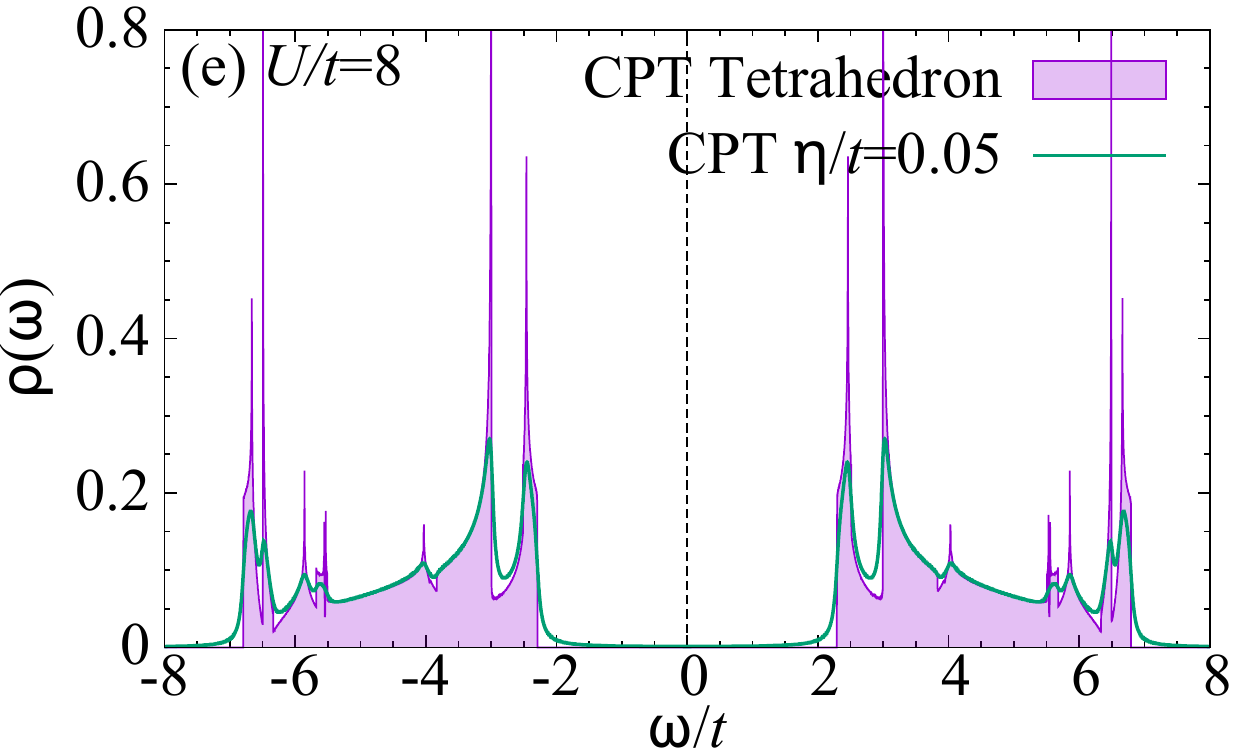}
    \caption{
      \label{fig.dos}
      Density of states $\rho(\w)$ (purple shaded regions) of the single-band Hubbard model 
      at $T=0$ for 
      (a) $U/t=1$, 
      (b) $U/t=2$, 
      (c) $U/t=4$, 
      (d) $U/t=6$, and 
      (e) $U/t=8$,       
      with $\mu=U/2$ 
      calculated using the tetrahedron method combined with the CPT. 
      A $2 \times 2$ site cluster is employed in the CPT. 
      For comparison, density of states $\rho_\eta(\w)$ calculated using 
      the standard procedure of the CPT with a finite Lorentzian broadening 
      $\eta/t=0.05$ is also shown by green solid lines. 
      The Fermi energy is indicated by dashed lines at $\w=0$. 
    }
  \end{center}
\end{figure}

Figure~\ref{fig.dos} shows the results of the density of states $\rho(\w)$ 
calculated using the tetrahedron method combined with the CPT. 
The calculations are done for several values of $U/t$ and $\mu=U/2$, i.e., at 
half-filling with the particle-hole symmetry. 
Since we consider a single-band model and a paramagnetic state, 
we simply drop the spin-orbital subscripts $\alpha$ and $\beta$ from the density of states 
$\rho_{\alpha\beta}(\w)$ in Eq.~(\ref{eq:tetdos}). 
For comparison, we also show in Fig.~\ref{fig.dos} the density of states calculated using 
the standard procedure of the CPT, i.e.,  
\begin{equation}
  \rho_\eta(\w) = 
  \frac{1}{V_{G}}\int_{\rm BZ} \dd^2 k \left[-\frac{1}{\pi}{\rm Im} G\left(\vec{\mb{k}},\w + \imag \eta \right) \right]  
  \label{eq:dos_cpt}
\end{equation} 
with a finite value of the Lorentzian broadening $\eta$ 
($\eta/t = 0.05$ in Fig.~\ref{fig.dos}) 
and the momentum integral is simply replaced by the sum of momenta discretized uniformly 
over the whole 2D Brillouin zone (also see Sec.~\ref{sec:part_BZ}). 
The number of discretized momenta taken in Eqs.~(\ref{eq:int_tetra}) and (\ref{eq:dos_cpt})
is $N_1 \times N_2 = 160  \times 160$ 
for all calculations shown in Fig.~\ref{fig.dos}. Note that the number $N_T$ of triangles introduced 
in the tetrahedron scheme is twice as large as this number. 
We find that the results are already converged 
at $N_1 \times N_2 = 100 \times 100$ for both tetrahedron and finite-broadening schemes. 

As shown in Fig.~\ref{fig.dos}, the overall structures of 
the density of states calculated with the finite 
Lorentzian broadening are in good accordance with those obtained 
by the tetrahedron method. 
However, the density of states calculated using the tetrahedron method 
is overwhelmingly sharp and the fine peak structures as well as 
small gaps can be easily distinguished. 
For example, the Mott gap is rather clearly observed in the density of 
states obtained by the tetrahedron method and can be evaluated 
even when the gap is tiny for small $U$.

Two remarks are in order. First, it is observed in Fig.~\ref{fig.dos} that there seem to exist 
spiky peaks in the density of states for all values of $U$. These seemingly 
spiky peaks are not delta-function peaks but have well defined Van-Hove 
like structures. These singularities appear 
where the stationary condition of the band dispersions is satisfied, i.e., 
$\nabla_{\vec{\mb{k}}}\w_m(\vec{\mb{k}})=0$. 
For example, each seemingly spiky peak in Fig.~\ref{fig.dos}(e) well corresponds to the stationary 
energy of the band dispersions shown in Fig.~\ref{fig.bands}(b). As described in Sec.~\ref{sec:lehmann} 
and also in appendix~\ref{appendix}, 
the many-body interactions are 
responsible for the emergence of these multiple energy bands (in principle, 
$N_{\rm pole}$ numbers of bands). 

Second, when the momentum $\veck_{\rm F}$ at which the gap opens is {\it a priori} known, 
one can readily calculate the single-particle gap simply by diagonalizing $\bs{M}(\veck_{\rm{F}})$. 
This is the case, for instance, 
for the half-filled Hubbard model on the one-dimensional chain ($\veck_{\rm F}=\pm\pi/2$) and 
on the honeycomb lattice ($\veck_{\rm F}$: $K$ and $K'$ points). 
It should be noted however that the 
periodization formula in Eq.~(\ref{eq.cpt})  
restores only the translational symmetry 
but not the point-group symmetry of the underlying lattice that is broken by the cluster partitioning. 
Therefore, the cluster partitioning should be carefully chosen 
in order not to break the point-group symmetry of the underlaying lattice. 
Otherwise, for instance, the Dirac points can deviate from the $K$ and $K'$ points 
in the CPT calculation for the honeycomb lattice, 
and therefore the simple diagonalization of $\bs{M}(\veck_{\rm F})$ 
is not enough to estimate the single-particle gap.

\section{Discussion}\label{sec:discussion}

In this section, we briefly discuss the extension of the tetrahedron method 
combined with the CPT 
to symmetry broken states, 3D systems, and finite-temperature calculations.

\subsection{Symmetry broken states}\label{sec_vca}

Using the variational cluster approximation (VCA)~\cite{Potthoff2003_PRL,Dahnken2004}, 
the present method can be adapted to symmetry broken states straightforwardly. 
The VCA introduces symmetry breaking Weiss fields $\mb{x}=(x_1,x_2,x_3,\dots)$ to the system 
and determines the optimal value $\mb{x}_{\rm op}$ which satisfies 
the stationary condition 
$\nabla_{\mb{x}} \Omega(\mb{x}) |_{\mb{x}=\mb{x}_{\rm op}} = \mb{0}$ for 
the grand-potential functional $\Omega(\mb{x})$~\cite{Potthoff2003,Potthoff2012}. 
Once the Weiss fields are optimized, we can easily obtain 
the single-particle Green's function 
$G_{\alpha \beta}(\vec{\mb{k}},z,\mb{x}_{\rm op})$ 
with $\mb{x}_{\rm op} \not = \bs{0}$~\cite{Dahnken2004}. 
Using this Green's function $G_{\alpha \beta}(\vec{\mb{k}},z,\mb{x}_{\rm op})$, 
the physical quantities including 
the density of states can be calculated for symmetry broken states. 
Technical details of the VCA can be found in 
Ref.~\cite{Senechal2008} for $T=0$ and 
Ref.~\cite{Eder2008} for finite temperatures. 
Since the Lehmann representation of the single-particle Green's function 
$G_{\alpha \beta}(\vec{\mb{k}},z,\mb{x}_{\rm op})$ is exactly the same form as in Eq.~(\ref{eq:Gab}), 
we can apply the tetrahedron method introduced in Sec.~\ref{sec:formalism} to symmetry broken states.

We should note here that in the VCA 
we also encounter the momentum integral for the grand-potential functional
over the whole Brillouin zone. 
Indeed, the grand-potential functional $\Omega$ per site 
in the zero-temperature limit is evaluated as  
\begin{eqnarray} \label{eq:grand}
  \Omega &=& \Omega'
   -  \frac{1}{L_{\rm c}}      \sum_{m=1}^{N_{\rm pole}} \lambda_m \Theta(E_{\rm F} - \lambda_m) \notag \\
  &+& \frac{1}{L_{\rm c} \tilde{V}_G} \sum_{m=1}^{N_{\rm pole}} \int_{\rm RBZ} \dd^2 \tilde{k} \w_m(\veckt) \Theta\left(E_{\rm F} - \w_m(\veckt)\right),  
\end{eqnarray} 
where $\Omega'$ is the exact grand potential per site of the cluster, 
$\Theta$ is the step function, 
$\veckt$ now denotes the momentum in the reduced Brillouin zone  
for the superlattice on which the clusters are defined, 
and $\tilde{V}_G$ is the volume of the reduced Brillouin zone~\cite{Aichhorn2006,Senechal2008}. 
Notice that in general $\lambda_m$ and $\w_m(\veckt)$ can depend on the spin-orbital index $\alpha$ 
in symmetry broken states and the dependencies on $\alpha$ are implicitly incorporated in $m$. 
Although the Fermi energy $E_{\rm F}$ is set to be zero in the preceding sections, 
here we show it explicitly for clarity. 

It is noticed in Eq.~(\ref{eq:grand}) that the third term on the right-hand side 
is the momentum integral for the effective single-particle energy $\w_m (\veckt)$ 
below the Fermi energy. 
We can therefore apply the tetrahedron method for 
the calculation of the grand potential functional in the VCA. 
For this purpose, the integration weight $w_{m,\tau,i}(\w)$ is required 
because the momentum integral involves the step function $\Theta$, instead of the 
delta function. 
Namely, we can evaluate the momentum integral in the tetrahedron method as  
\begin{eqnarray} 
  &&\frac{1}{L_{\rm c} \tilde{V}_G} \int_{\rm RBZ} \dd^2 \tilde{k}  \w_m(\veckt) \Theta\left(E_{\rm F} - \w_m(\veckt)\right) \notag \\
  &&\quad\quad\quad  = \frac{1}{L_{\rm c}} \sum_{\tau=1}^{N_\tau}  \sum_{i=1}^{3} w_{m,\tau,i} (E_{\rm F}) \w_m(\veckt_i),  
\end{eqnarray} 
where the triangular partitioning should be applied for the reduced Brillouin zone and the integration 
weight $w_{m,\tau,i}(\w)$ is provided in Table~\ref{table}.

\subsection{3D systems} 

Although the formulation provided in Sec.~\ref{sec:formalism} 
is for 2D interacting fermion systems, the application to 3D systems 
is straightforward. Simply referring to the original papers on 
the tetrahedron method in the single-particle theory~\cite{Rath1975,Blochl1994},  
one can easily complete Table~\ref{table} for 3D interacting fermion systems. 
For example, the VCA has been employed recently to investigate 
the competing magnetic orders and single-particle excitations 
of the single-band Hubbard model on 
the stacked square lattice~\cite{Yoshikawa2009} and 
the simple cubic lattice~\cite{Laubach2016}
with longer-range hoppings. 

\subsection{Finite temperatures} \label{sec:ft}

The tetrahedron method introduced in Sec.~\ref{sec:formalism} 
requires diagonalizing 
the effective single-particle Hamiltonian matrix  
$\bs{M}\ofk$ at each momentum $\vec{\mb{k}}$
to obtain the single-particle excitation energies and spectral-weight functions for 
interacting fermions. 
At zero temperature, the dimension of $\bs{M}(\vec{\mb{k}})$  
is typically $N_{\rm pole} \sim \mcal{O}(10^2)$ and thus the numerical 
diagonalization of $\bs{M}(\vec{\mb{k}})$ is computationally not expensive. 
On the other hand, at finite temperatures,
$N_{\rm pole}$ reaches $ \mcal{O}(10^3)$ and can become much larger~\cite{Aichhorn2003,Eder2008,Eder2010chi}. 
This is certainly the case even at zero temperature when the size of clusters is large. 
However, $\bs{M}(\vec{\mb{k}})$ can be independently diagonalized at 
different momenta and 
thus the $\vec{\mb{k}}$-point parallelization is highly efficient to reduce the elapsed time.

\section{Summary}\label{sec:summary}

By combining the tetrahedron method with the CPT, 
we have introduced a method to numerically calculate the density of states of interacting fermions. 
The method removes the Lorentzian broadening parameter $\eta$ to represent delta-function peaks 
and thus allows us to resolve fine structures of the density of states and evaluate a single-particle excitation gap 
without performing the extrapolation of $\eta \to 0$. 
The formulation is based on the Lehmann representation of the interacting 
single-particle Green's function in the CPT. 
We have emphasized the notion of the effective single-particle Hamiltonian 
for the single-particle excitation energies of interacting fermions, 
which is the conceptual basis of the proposed method and enables us to apply the tetrahedron 
method developed in the single-particle theory. 
The general correspondence between the self-energy and the effective single-particle Hamiltonian 
has also been established. 

The formalism has been provided in detail for 2D multi-orbital interacting systems
and the benchmark calculation has been performed for the 2D single-band Hubbard model. 
The method can be easily adapted to symmetry broken states using the VCA. 
We have also argued that extension of the formalism to 3D interacting systems is straightforward. 
For the finite-temperature calculation, 
the $\vec{\mb{k}}$-point parallelization to diagonalize the 
effective single-particle Hamiltonian $\bs{M}(\vec{\mb{k}})$ 
would be required because of the rapid increase of the dimension $N_{\rm pole}$ of 
$\bs{M}(\vec{\mb{k}})$ with the temperature.  
Not only the method itself but also the notion of 
the effective single-particle Hamiltonian is a useful concept 
to investigate the single-particle excitations of interacting fermions in general.

\acknowledgments 
The computations have been done using the RIKEN Integrated Cluster of Clusters (RICC) facility and 
the RIKEN supercomputer system (HOKUSAI GreatWave). 
This work has been supported in part by 
Grant-in-Aid for Scientific Research from MEXT Japan 
under the Grant No. 25287096, and also by 
RIKEN iTHES Project and Molecular Systems. 

\appendix
\section{Effective single-particle Hamiltonian and self-energy}\label{appendix}

In this appendix, 
we first construct an effective single-particle Hamiltonian whose eigenvalues 
coincide with the single-particle excitation energies of interacting fermions.  
Next, we establish the general correspondence between the self-energy and the matrix elements of 
this effective single-particle Hamiltonian. 
It should be emphasized that the formalism developed in this appendix 
is not specific to the CPT but relevant for any interacting fermions.

To simplify the notation, here we only consider a single-orbital system 
in a paramagnetic state. 
Therefore, the single-particle Green's function is 
independent of the spin directions, i.e., 
$G_{\up \up}\ofkz = G_{\dn \dn}\ofkz \equiv G\ofkz$ and 
$G_{\up \dn}\ofkz = G_{\dn \up}\ofkz = 0$. 
However, the extension to a multi-orbital system 
is straightforward~\cite{Seki2016}.

From the Dyson equation, the inverse of the single-particle Green's function $G\ofkz$
is given as 
\begin{eqnarray}
  G^{-1}\ofkz
  &=& G_{0}^{-1}\ofkz - \Sigma\ofkz \notag \\
  &=& z - \eps_{\vec{\mb{k}}} - \Sigma_{0} (\vec{\mb k})- \sum_{\nu=1}^{N_{\rm zero}} 
  \frac{|\Delta_{\vec{\mb{k}} \nu}|^2}{z-\zeta_{\vec{\mb{k}}  \nu}},  
  \label{eq.QPGreen}
\end{eqnarray}
where $\eps_{\vec{\mb{k}}}$ and $G_{0}\ofkz = (z - \eps_{\vec{\mb{k}}})^{-1}$ are the single-particle energy and 
the single-particle Green's function 
in the noninteracting limit, respectively. 
In the last equality of Eq.~(\ref{eq.QPGreen}), the self-energy $\Sigma\ofkz$ 
is written in the Lehmann 
representation and is divided into a real static part $\Sigma_0(\vec{\mb k})$ (such as the Hartree potential) and 
a sum of poles $\zeta_{\vec{\mb{k}} \nu}$ 
on the real frequency axis~\cite{Luttinger1961,Eder2007,Eder2008,Eder2011,Dzyaloshinskii2003}. 
Notice that $N_{\rm zero}$ is the number of poles of the self-energy $\Sigma\ofkz$ and is 
exactly the same as the number of zeros of the single-particle Green's function $G\ofkz$~\cite{Eder2014}. 
Therefore, $N_{\rm zero}=N_{\rm pole}-1$, where $N_{\rm pole}$ is the number of poles of 
the single-particle Green's function $G\ofkz$. 
It should also be noted that Eq.~(\ref{eq.QPGreen}) is valid at any temperature and the temperature dependence of 
$G\ofkz$ and $\Sigma\ofkz$ is implicitly assumed. 

We define the following single-particle Hermitian operator $\hat{h}$: 
\begin{eqnarray}
  \hat{h} 
  &=& 
  \sum_{\vec{\mb{k}}} 
  \hat{\mb{c}}_{\vec{\mb{k}}  }^\dag 
  \bs{h}_{\vec{\mb{k}} } 
  \hat{\mb{c}}_{\vec{\mb{k}} }, 
  \label{eq:heff}
\end{eqnarray}
where 
\begin{eqnarray}
  \hat{\mb{c}}_{\vec{\mb{k}}}^\dag
  &=& 
  \left(
    \begin{array}{cccc}    
      \hat{c}_{\vec{\mb{k}}}^\dag,  &
      \hat{x}_{\vec{\mb{k}} 1}^\dag, &
      \cdots, &
      \hat{x}_{\vec{\mb{k}} N_{\rm zero}}^\dag 
    \end{array} 
  \right) 
  \label{eq:ceff}
\end{eqnarray}
is a set of fermion creation operators and  
\begin{eqnarray}
  \bs{h}_{\vec{\mb{k}}}&=&
  \left[
    \begin{array}{c|cccc}
      \eps_{\vec{\mb{k}}} + \Sigma_{0}(\vec{\mb k}) & \Delta_{\vec{\mb{k}} 1}^* & \Delta_{\vec{\mb{k}} 2}^* & \cdots & \Delta_{\vec{\mb{k}} N_{\rm zero}}^* \\
      \hline
      \Delta_{\vec{\mb{k}} 1} & \zeta_{\vec{\mb{k}} 1} & 0 & \cdots & 0 \\
      \Delta_{\vec{\mb{k}} 2} & 0 & \zeta_{\vec{\mb{k}} 2} & \ddots & \vdots \\
      \vdots              & \vdots  & \ddots       & \ddots & 0 \\
      \Delta_{\vec{\mb{k}} N_{\rm zero}} & 0 & \cdots & 0 & \zeta_{\vec{\mb{k}} N_{\rm zero}} 
    \end{array}
    \right]\label{eq:hmat1}  \\
  &=&
  \left[
    \begin{array}{c|c}
      \eps_{\vec{\mb{k}}} + \Sigma_{0}(\vec{\mb k}) & \bs{\Delta}_{\vec{\mb{k}}}^\dag \\ 
      \hline
      \bs{\Delta}_{\vec{\mb{k}}} & \bs{\zeta}_{\vec{\mb{k}}} \\
    \end{array}
    \right] \label{eq:hmat}
\end{eqnarray} 
is the $(N_{\rm zero}+1)\times(N_{\rm zero}+1)$ Hermitian matrix~\cite{arrowhead}. 
In the above matrix representation of $\hat h$, the horizontal and vertical lines are indicated 
to distinguish the real and auxiliary fermion spaces (see below). 
We have also introduced in Eq.~(\ref{eq:hmat}) that the $N_{\rm zero}$ dimensional row vector 
\begin{equation}
\bs{\Delta}_{\vec{\mb{k}}}^\dag
=(\Delta_{\vec{\mb{k}} 1}^*, \Delta_{\vec{\mb{k}} 2}^*, \dots, \Delta_{\vec{\mb{k}} N_{\rm zero}}^*)
\end{equation} 
and the $N_{\rm zero}\times N_{\rm zero}$ diagonal matrix 
\begin{equation}
\bs{\zeta}_{\vec{\mb{k}}} = {\rm diag}
(\zeta_{\vec{\mb{k}} 1}, \zeta_{\vec{\mb{k}} 2}, \cdots ,\zeta_{\vec{\mb{k}} N_{\rm zero}}).
\end{equation}

We can now easily show that 
the first diagonal component of $(z-\bs{h}_{\vec{\mb{k}}})^{-1}$ is the single-particle Green's 
function $G\ofkz$ given in Eq.~(\ref{eq.QPGreen}). 
Since the Schur's complement of the lower right block $z - \bs{\zeta}_{\vec{\mb{k}}}$ of 
$z-\bs{h}_{\vec{\mb{k}}}$ is 
\begin{equation} 
  \widetilde{(z-\bs{h}_{\vec{\mb{k}}})}_{11}
  = z - \eps_{\mb k } - \Sigma_{0}(\vec{\mb k}) - \bs{\Delta}_{\vec{\mb{k}}}^\dag (z - \bs{\zeta}_{\vec{\mb{k}}})^{-1} \bs{\Delta}_{\vec{\mb{k}}}, \label{eq.Schur}
\end{equation}
we indeed find that 
\begin{equation}
\widetilde{(z-\bs{h}_{\vec{\mb{k}}})}_{11} = G^{-1}\ofkz.
\end{equation}
Here $z \not = \zeta_{\vec{\mb{k}}\nu}$ is assumed in order that $z - \bs{\zeta}_{\vec{\mb{k}}}$ is regular. 
Therefore, assuming that $\bs{h}_{\vec{\mb{k}}}$ is diagonalized by a unitary matrix $\bs{u}_{\vec{\mb{k}}}$ 
as 
\begin{equation}
\bs{u}_{\vec{\mb{k}}}^\dag \bs{h}_{\vec{\mb{k}}} \bs{u}_{\vec{\mb{k}}} 
= {\rm diag} (\w_{\vec{\mb{k}}1}, \cdots, \w_{\vec{\mb{k}}N_{\rm pole}}),
\label{eq:h_diag}
\end{equation}  
we find that 
\begin{equation}
  G\ofkz = \sum_{m=1}^{N_{\rm pole}} \frac{\left |[\bs{u}_{\vec{\mb{k}}}]_{m1}\right |^2}{z-\w_{\vec{\mb{k}}m}},
  \label{eq:guw}
\end{equation} 
indicating that the spectral-weight function for the $m$-th pole is simply given by 
$\left |[\bs{u}_{\vec{\mb{k}}}]_{m1}\right |^2$.

The eigenvalues of $\bs{h}_{\vec{\mb{k}}}$ are given as the roots of secular equation 
$\det{(z - \bs{h}_{\vec{\mb{k}}})}=0$. 
From the block matrix determinant formula 
\begin{eqnarray}
  \det\left[
    \begin{array}{c|c}
      \bs{A} & \bs{B} \\ 
      \hline
      \bs{C} & \bs{D} 
    \end{array}
    \right] 
  = \det \bs{D} \cdot \det{\left(\bs{A} - \bs{B}\bs{D}^{-1}\bs{C} \right)}, 
\end{eqnarray}
we find that 
\begin{eqnarray}
  \det{(z - \bs{h}_{\vec{\mb{k}}})} 
  &=& \det{(z - \bs{\zeta}_{\vec{\mb{k}}})} \det{\widetilde{(z - \bs{h}_{\vec{\mb{k}}})}_{11}} \notag \\
  &=& \det{(z - \bs{\zeta}_{\vec{\mb{k}}})} \det{G\ofkz^{-1}} \label{detz_h}
\end{eqnarray} 
and thus the eigenvalues of $\bs{h}_{\vec{\mb{k}}}$ coincide with the poles of $G\ofkz$, 
as explicitly shown in Eq.~(\ref{eq:guw}). 
Note that $\det{G\ofkz} = G\ofkz $ since here we consider the single-band 
paramagnetic system.  
Therefore, the single-particle Hamiltonian $\hat{h}$ introduced above can be considered 
as an effective single-particle Hamiltonian 
for the single-particle excitation energies of interacting fermions where the 
single-particle Green's function $G\ofkz$ is given in Eq.~(\ref{eq.QPGreen}).

In other words, the single-particle excitations of interacting fermions are described exactly 
by the fermionic single-particle Hamiltonian $\hat{h}$, where 
the real fermions $\hat{c}_{\vec{\mb{k}}}^\dag$ 
hybridizes to the ``auxiliary'' fermions 
$\hat{x}_{\vec{\mb{k}}\nu}^\dag$ ($\nu=1,2,\dots,N_{\rm zero}$) with the hybridization $\Delta_{\vec{\mb{k}}\nu}$ 
and the energy dispersions of the auxiliary fermions $\hat{x}_{\vec{\mb{k}}\nu}^\dag$ are given 
by the poles $\zeta_{\vec{\mb{k}}\nu}$ of the self-energy $\Sigma\ofkz$. 
It is interesting to remark that 
not only the low-energy single-particle excitations, 
e.g., quasiparticles in the Landau's Fermi liquid theory~\cite{Landau1956,Nozieres1962,Luttinger1962},   
but also high-energy single-particle excitations (even in, e.g.,  a Mott insulating state) 
are formally described by the fermionic single-particle Hamiltonian.

Equation~(\ref{detz_h}) indicates that 
the single-particle Green's function $G\ofkz$ can be written in the rational polynomial form 
\begin{equation}\label{eq.ratio}
  G\ofkz = 
  \frac{\det{(z - \bs{\zeta}_{\vec{\mb{k}}})}}{\det{(z - \bs{h}_{\vec{\mb{k}}})}} = 
  \frac{\prod_{\nu=1}^{N_{\rm zero}} (z - \zeta_{\vec{\mb{k}}\nu}) }{\prod_{m=1}^{N_{\rm pole}} (z - \w_{\vec{\mb{k}} m}) },  
\end{equation} 
where $\w_{\vec{\mb{k}} m}$ are the eigenvalues of $\bs{h}_{\vec{\mb{k}}}$ [see Eq.~(\ref{eq:h_diag})]. 
Equation~(\ref{eq.ratio}) explicitly shows that the zeros of $G\ofkz$ corresponds to the poles of $\Sigma \ofkz$. 
The asymptotical behaviour of $G\ofkz \sim 1/z$ for large $|z|$ is also apparent 
from Eq.~(\ref{eq.ratio}) because $N_{\rm zero}=N_{\rm pole} - 1$, which 
ensures the spectral-weight sum rule or equivalently the correct zero-th moment of the single-particle 
Green's function $G\ofkz$.

Although we have shown the correspondence between $\bs{h}_{\vec{\mb{k}}}$ and $\Sigma\ofkz$, 
the ``single-particle" parameters $\Sigma_0(\vec{\mb k})$, $\bs{\Delta}_{\vec{\mb{k}} \nu}$, 
and $\bs{\zeta}_{\vec{\mb{k}}\nu}$ in $\bs{h}_{\vec{\mb{k}}}$ are still unknown.  
In principle, these parameters can be extracted from the numerically calculated self-energy 
of interacting fermions, as, for example, in Ref.~\cite{Eder2011}. 
Alternatively, we can impose constraints on these parameters 
by considering the moment of the single-particle Green's function $G\ofkz$~\cite{Harris1967}. 
For example, 
the second-order moment of the single-particle Green's function, 
which guarantees the spectral-weight sum rule of the self-energy, 
imposes that 
\begin{equation}
  \bs{\Delta}_{\vec{\mb{k}} }^\dag \bs{\Delta}_{\vec{\mb{k}}} 
  =  U^2 n (1-n)  
\end{equation}
for the single-band Hubbard model in a paramagnetic state, 
where $U$ is the on-site Coulomb repulsion and $n$ is the electron density per spin~\cite{Seki2011}. 
The higher-order moments of the single-particle Green's function 
may impose further constraints for $\bs{\Delta}_{\vec{\mb{k}}}$ 
and other parameters~\cite{Turkowski2008}.

Instead of extracting the ``single-particle" parameters from the self-energy $\Sigma \ofkz$, 
the CPT derives the effective single-particle Hamiltonian $\bs{M}\ofk$  
from $\bs{Q}$, $\bs{\Lambda}$, and $\bs{\mcal{T}}\ofk$, i.e., 
the exact self-energy of the cluster~\cite{Potthoff2003_PRL,Dahnken2004,Gros1993,Senechal2012}. 
However, we should note that generally $\bs{M}\ofk$ is different from $\bs{h}_{\vec{\mb{k}}}$. 
This is simply because the CPT is an approximation scheme to obtain the single-particle Green's function 
$G\ofkz$ of interaction fermions and thus $\bs{M}\ofk$ is considered as an approximation of 
$\bs{h}_{\vec{\mb{k}}}$. 

On the analogy of the mean-field theory, 
the spectral weight $\Delta_{\vec{\mb{k}}\nu}$ of the self-energy $\Sigma \ofkz$ 
plays the role of ``gap function'' generated 
by fermion correlations without breaking any symmetry. 
The $\vec{\mb{k}}$ independent (dependent) $\Delta_{\vec{\mb{k}}\nu}$ hence implies that  
the effect of fermion correlations 
is local (non-local). 
In this regard, the presence of such a ``gap function'' with $d_{x^2-y^2}$-wave symmetry   
has been reported 
in an exact diagonalization study for the $t$-$J$ model 
on a small cluster~\cite{Ohta1994,Poilblanc2002}. 
As described above, the ``gap function'' $\Delta_{\vec{\mb{k}}\nu}$ represents the hybridization between 
the real fermion and the auxiliary fermion $\hat{x}_{\vec{\mb{k}}\nu}$, 
which appears in the form of $\hat{c}_{\vec{\mb{k}}}^\dag \hat{x}_{\vec{\mb{k}}\nu}$. 
Therefore, if the energy dispersion $\zeta_{\vec{\mb{k}}\nu}$ of the auxiliary fermion 
is found approximately as 
$\zeta_{\vec{\mb{k}}\nu} \simeq  \eps_{\vec{\mb{k}}+\vec{\mb{Q}}}$ 
$\ $($\zeta_{\vec{\mb{k}}\nu} \simeq -\eps_{\vec{\mb{k}}}$), 
then $\hat{x}_{\vec{\mb{k}},\nu}$ would be characterized dominantly by 
$\hat{c}_{\vec{\mb{k}}+\vec{\mb{Q}}}$ ($\hat{c}^\dag_{\vec{\mb{k}}}$), 
indicating the presence of excitonic (superconducting) type fluctuations 
induced by fermion correlations. 

Finally, we note that the single-particle Hamiltonians considered in Refs.~\cite{Sakai2014,Sakai2015} 
to describe the single-particle excitations obtained by 
the cellular dynamical mean field theory (CDMFT) can be considered as 
the single-particle Hamiltonian $\hat{h}_{\vec{\mb{k}}}$ in Eq.~(\ref{eq:heff}) with the 
``single-particle" parameters $\zeta_{\vec{\mb{k}}\nu}$ and $\Delta_{\vec{\mb{k}} \nu}$ 
extracted from the CDMFT calculations. 
The hidden fermions in Refs.~\cite{Imada2011,Sakai2014,Sakai2015} correspond to the 
auxiliary fermions described by $\hat{x}^\dag_{\vec{\mb{k}} \nu}$ in Eq.~(\ref{eq:ceff}).

\section{Calculation of the density of states in the reduced Brillouin zone}\label{appendixB}

In this appendix, we show that within the CPT the density of states can also be calculated 
by integrating the imaginary part of $\tilde{\bs{G}}\ofkz$ over the reduced Brillouin zone 
for the superlattice on which the clusters are defined. 
As shown in the following, this is due to the fact that the periodization formula in 
Eq.~(\ref{eq.cpt}) does not change the trace of the single-particle Green's function~\cite{Senechal2012}, i.e., 
$
\frac{1}{V_G}\int_{\rm BZ} \dd^2k G_{\alpha \beta}\ofkz = 
\frac{1}{L_{\rm c} \tilde{V}_G}\int_{\rm RBZ} \dd^2\tilde{k}
\sum_{i=1}^{L_{\rm c}} \tilde{G}_{i\alpha,i\beta}(\veckt,z)$.
 
For this purpose, we consider the local single-particle Green's function  
\begin{equation}\label{eq:g_local}
  G_{\alpha \beta}(z) = \frac{1}{V_{G}} \int_{\rm BZ} \dd^2 k G_{\alpha \beta} \ofkz,  
\end{equation} 
where $G_{\alpha \beta} \ofkz$ is the single-particle Green's function of $\hat H$ on the infinite lattice 
given in Eq.~(\ref{eq.cpt}). 
Following Refs.~\cite{Senechal2002,Senechal2012,Senechal2008}, 
we represent a momentum $\veck$ as 
\begin{equation}
  \veck = \vecqs + \veckt, 
\end{equation}
where $\vecqs$ is a reciprocal lattice vector of the superlattice, 
specifying the location of the $s$-th reduced Brillouin zone
($s=1,2,\cdots,L_{\rm c}$), and 
$\veckt$ is a momentum in the first reduced Brillouin zone. 
Since $\exp{(\imag \vecqs \cdot \vec{\bs{R}}_I)} = 1$ by definition, 
it follows from Eqs.~(\ref{eq.Gcpt}) and (\ref{Vij}) that 
$\mcal{T}(\veck) = \mcal{T}(\veckt)$ and $\tilde{\bs{G}}(\veck,z)=\tilde{\bs{G}}(\veckt,z)$. 
Therefore, the momentum integral over the Brillouin zone in Eq.~(\ref{eq:g_local}) can be 
reduced into the momentum integral over the first reduced Brillouin zones for the superlattice, i.e.,
\begin{widetext}
\begin{eqnarray}
  G_{\alpha \beta}(z) 
  &=&
  \sum_{s=1}^{L_{\rm c}} \frac{1}{L_{\rm c} \tilde{V}_{G}} \int_{{\rm RBZ}_s} \dd^2 \tilde{k} G_{\alpha \beta} \ofkz \nonumber \\
    &=& 
  \sum_{s=1}^{L_{\rm c}} \frac{1}{L_{\rm c}^2 \tilde{V}_{G}} \int_{{\rm RBZ}_s} \dd^2 \tilde{k} 
  \sum_{i=1}^{L_{\rm c}} \sum_{j=1}^{L_{\rm c}} 
  \tilde{G}_{i\alpha,j\beta}(\veckt,z) 
  \e^{-\imag \veckt \cdot(\vec{\bs{r}}_i - \vec{\bs{r}}_j)} 
  \e^{-\imag \vecqs \cdot (\vec{\bs{r}}_i - \vec{\bs{r}}_j)} \notag \\
  &=&
  \frac{1}{L_{\rm c}\tilde{V}_{G}} \int_{\rm RBZ} \dd^2 \tilde{k} 
  \sum_{i=1}^{L_{\rm c}} 
  \tilde{G}_{i\alpha,i\beta}(\veckt,z),  
\end{eqnarray}
\end{widetext}
where the momentum integral $\int_{{\rm RBZ}_s} \dd^2 \tilde{k}\cdots$ in the first and second equations 
is over the $s$-th reduced 
Brillouin zone, while the momentum integral $\int_{\rm RBZ} \dd^2 \tilde{k}\cdots$ in the last equation is over 
the first reduced Brillouin zone for the superlattice. 
In the last equality, we have used that 
\begin{equation} 
  \frac{1}{L_{\rm c}}
  \sum_{s=1}^{L_{\rm c}} \e^{-\imag \vecqs \cdot(\vec{\bs{r}}_i - \vec{\bs{r}}_j)} =
  \delta_{\vec{\bs{r}}_i,\vec{\bs{r}}_j},
\end{equation} 
where we employ the convention given in Eq.~(\ref{eq:position}) to represent the position of 
a site in the infinite lattice~\cite{Senechal2008}. 

Since the density of states $\rho_{\alpha \beta} (\w)$ is evaluated as 
\begin{equation}
  \rho_{\alpha \beta} (\w)= -\frac{1}{\pi} \lim_{\eta \to 0} {\rm Im} G_{\alpha \beta} (\w + \imag \eta),
\end{equation} 
we find that 
\begin{equation}
  \rho_{\alpha \beta} (\w) 
  =\sum_{m=1}^{N_{\rm pole}} \frac{1}{\tilde{V}_{G}}\int_{\rm RBZ} \dd^2 \tilde{k} \tilde{A}_{\alpha \beta,m}(\vec{\tilde{\mb{k}}}) \delta\left(\w - \w_m(\vec{\tilde{\mb{k}}})\right), 
  \label{eq:dos2}
\end{equation}
where the spectral-weight function for the reduced Brillouin zone is given as 
\begin{equation}
  \tilde{A}_{\alpha \beta, m}(\veckt) =  \frac{1}{L_{\rm c}}
    \sum_{i=1}^{L_{\rm c}} \tilde{Q}_{i \alpha, m}(\veckt) \tilde{Q}_{i \beta, m}^*(\veckt).  
  \label{eq.At}
\end{equation}
We can therefore apply the tetrahedron method combined with the CPT in the reduced Brillouin zone 
simply by replacing the spectral-weight function $A_{\alpha \beta, m}(\vec{\mb k})$ in Eq.~(\ref{eq:dos}) 
with the modified spectral-weight function $\tilde{A}_{\alpha \beta, m}(\veckt)$ given in Eq.~(\ref{eq.At}).

Since the area of the reduced Brillouin zone is $L_{\rm c}$ times smaller than the original one, 
the momentum mesh in the reduced Brillouin zone can be $L_{\rm c}$ times denser than 
that in the whole Brillouin zone when the total number of meshes is the same. 
In other words, adopting the momentum integral over the reduced Brillouin zone, we can save the 
computational effort by a factor of $L_{\rm c}$ 
to obtain the density of states with the same accuracy. 
In addition, 
$\tilde{A}_{\alpha \beta,m}(\veckt)$ is rather easily calculated than $A_{\alpha \beta,m}(\vec{\mb k})$.


\begin{thebibliography}{99}
\bibitem{arrowhead}
  The sparse matrix of the form in Eq.~(\ref{eq:hmat1}), 
  which contains non-zero matrix elements only in the first column, first row, and diagonal, 
  is known as arrowhead matrix.
\end{thebibliography}
\end{document}